\documentclass[11pt,a4paper]{article}
\usepackage{jcappub}

\usepackage{bm}
\usepackage{url}
\usepackage{doi}
\usepackage{multirow}
\usepackage{epstopdf}
\usepackage{subfigure}

\def\gtsima{$\; \buildrel > \over \sim \;$}
\def\ltsima{$\; \buildrel < \over \sim \;$}
\def\gsim{\lower.5ex\hbox{\gtsima}}
\def\lsim{\lower.5ex\hbox{\ltsima}}
\def\simleq{\; \raise0.3ex\hbox{$<$\kern-0.75em \raise-1.1ex\hbox{$\sim$}}\; }

\newcommand{\GeV}{{\rm GeV}}

\newcommand{\TeV}{{\rm TeV}}

\newcommand{\kpc}{{\rm kpc}}

\newcommand{\cm}{{\rm cm}}
\newcommand{\km}{{\rm km}}
\newcommand{\muG}{\mu{\rm G}}

\newcommand{\s}{{\rm s}}

\newcommand{\dragon}{{\sffamily DRAGON}}

\begin{document}


\title{Cosmic Ray Electrons, Positrons and the Synchrotron emission of the Galaxy: consistent analysis and implications}
\author[a,b]{Giuseppe Di Bernardo}
\author[c]{Carmelo Evoli}
\author[d,e]{Daniele Gaggero}
\author[f,g]{Dario Grasso}
\author[h,i]{Luca Maccione}

\affiliation[a]{Department of Physics, University of Gothenburg, SE 412 96 Gothenburg, Sweden}
\affiliation[b]{Department of Astronomy and Theoretical Astrophysics Center, University of California Berkeley, Berkeley, CA 94720}
\affiliation[c]{{II.} Institut f\"ur Theoretische Physik, Universit\"at Hamburg, Luruper Chaussee 149, 22761 Hamburg, Germany}
\affiliation[d]{SISSA, Via Bonomea 265, 34136 Trieste, Italy}
\affiliation[e]{Institut fŸr Experimentelle Kernphysik, Karlsruhe Institute of Technology, P.O. Box 6980, 76128 Karlsruhe, Germany}
\affiliation[f]{Dipartimento di Fisica, Universit\`a di Siena, Via Roma 56, I-56100 Siena, Italy}
\affiliation[g] {Istituto Nazionale di Fisica Nucleare, Sezione di Pisa, Largo B. Pontecorvo, I-56127, Pisa, Italy}
\affiliation[h]{Max-Planck-Institut f\"ur Physik (Werner-Heisenberg-Institut),
F\"ohringer Ring 6, D-80805 M\"unchen}
\affiliation[i]{Ludwig-Maximilians-Universit\"at, Arnold Sommerfeld Center, Theresienstra{\ss}e 37, D-80333 M\"unchen}

\emailAdd{giuseppe.dibernardo@physics.gu.se}
\emailAdd{carmelo.evoli@desy.de}
\emailAdd{dgaggero@sissa.it}
\emailAdd{dario.grasso@pi.infn.it}
\emailAdd{luca.maccione@lmu.de}

\date{\today}

\abstract{A multichannel analysis of cosmic ray electron and positron spectra and of the diffuse synchrotron emission of the Galaxy is performed by using the \dragon\ code. This study is aimed at probing the interstellar electron source spectrum down to $E \lesssim 1~\GeV$ and at constraining several propagation parameters. 
We find that above $4~\GeV$ the $e^-$ source spectrum is compatible with a power-law of index $\sim 2.5$. Below $4~\GeV$ instead it must be significantly suppressed and the total lepton spectrum is dominated by secondary particles.
The positron spectrum and fraction measured below a few GeV are consistently reproduced only within low reacceleration models. 
We also constrain the scale-height $z_t$ of the cosmic-ray distribution using three independent (and, in two cases, original) arguments, showing that values of $z_t \simleq 2~\kpc$ are excluded. This result may have strong implications for particle dark matter searches.}

\subheader{LMU-ASC 72/12; MPP-2012-135}
\maketitle
\flushbottom

\section{Introduction}

The latest few years have led to remarkable progress in the knowledge of the leptonic component of Galactic Cosmic Rays (CR). 
This was achieved mainly thanks to a set of successful experiments which measured the absolute, combined and relative spectra of electrons and positrons in space or in the high atmosphere (see \cite{Daniele} for a recent review). 

Among other relevant results, the PAMELA orbital observatory measured the positron fraction 
between 1 and 100 GeV \cite{Adriani:2008zr} finding that it unexpectedly increases above 10 GeV; Fermi-LAT determined the $(e^+ + e^-)$ spectrum in the energy range $7~ \GeV < E < 1~\TeV$ \cite{Abdo:2009zk,Ackermann:2010ij} to be compatible with a power-law of index $- 3.08 \pm 0.05$ and to display significant evidence of a spectral hardening above a few hundred GeV; PAMELA recently measured the $e^-$ spectrum between 1 and 625 GeV \cite{Adriani:2011xv} finding that above $30~\GeV$ it is described by a power-law of index $ - 3.18 \pm 0.05$ and that at lower energies it is almost in agreement with that previously measured by ASM-01 \cite{Aguilar:2007yf}. Furthermore the H.E.S.S. atmospheric Cherenkov telescope observed a pronounced steepening of the $(e^+ + e^-)$ spectrum above 1 TeV \cite{Aharonian:2009ah}.

All these results can consistently be described in the framework of semi-analytical or numerical propagation models accounting for the following spectral components: 
\begin{itemize}
\item standard primary electrons ($e^-$) originated by astrophysical sources, most likely Supernova Remnants (SNRs) ;
\item secondary electrons and positrons produced by the scattering of the nuclear component of CR with the ISM; 
\item a new electron + positron ($e^\pm$) primary component, of still unknown origin, taking over at about a few hundred GeV.  
\end{itemize}
The $e^\pm$ \emph{extra-component} is required in order to provide a consistent description of the rising positron fraction found by PAMELA and of the hard $(e^+ + e^-)$  spectrum measured by Fermi-LAT (see e.g.~\cite{Grasso:2009ma} and Ref.s therein).  Its presence has been recently confirmed by the measurement of the $e^+$ and $e^-$ separate spectra performed by Fermi-LAT \cite{FermiLAT:2011ab,Grasso:2011wt}. 

In spite of these successes several important questions on the origin and the propagation of cosmic-ray electrons (CRE) remain open. Due to several degeneracies among the relevant parameters, the rigidity dependence of the diffusion coefficient, as well as the role of reacceleration and convection in CR propagation are poorly known. 

Another very important source of uncertainty is the large error on the vertical (perpendicular to the Galactic plane) extension of the CR diffusion region $(1 \lesssim L \lesssim 10~\kpc)$ as estimated from the $^{10}{\rm Be}/ ^{9}$Be ratio \cite{Strong:2007nh}. Interestingly, recent results derived from a global analysis of the available data in the framework of numerical propagation yielded a stronger constraint $L = 5.4 \pm 1.4~\kpc$ (corresponding to a 95\% posterior range 3.2--8.6 kpc) \cite{Trotta:2010mx}, while the same kind of analysis performed with semi-analytic codes found much weaker constraints $L = 8 ^{+8}_{-7}~\kpc$ \cite{Putze:2010zn}. We do not discuss here the possible origin of such differences, but just remark that the theoretical systematics associated with the determination of this parameter may well be larger than the errors estimated by a statistical analysis. Therefore, it is important to use also other, possibly non-local, observables to constrain the scale-height $L$. These issues are relevant not only for conventional CR physics but also for dark matter (DM) indirect searches since the local flux of DM decay/annihilation products is expected to depend significantly on $L$ (see \cite{Evoli:2011id} for a recent analysis).  

Furthermore, we know very little about the spatial distribution of CRE sources. In particular, one of the most debated questions is whether the source(s) of the $e^\pm$ extra-component is local (e.g.~a nearby pulsar) or distributed in a large halo (as expected if its source is DM).   

Some of the issues mentioned above can hardly be solved using electrically charged messengers alone, because {\it a)} solar wind reshapes the charged CR spectra in a poorly known way, thereby preventing from probing their local interstellar spectra (LIS) at rigidities below a few tens of $\GeV$; 
{\it b)}  the diffusion of charged CRs in the random component of the galactic magnetic field (GMF) erases most of the angular information about their origin and spatial distribution in the Galaxy. 
 
The $\gamma$-ray and the synchrotron diffuse emissions of the Galaxy offer valuable complementary probes of the low energy spectrum and of the spatial distribution of CRs in the Galaxy. However, the $\gamma$-ray diffuse emission receives a significant contribution from proton-proton interactions in the interstellar medium. Therefore, we limit ourselves to consider the synchrotron emission as it offers a more direct probe of the leptonic CR component. We will use the \dragon\ numerical CR propagation code to consistently model the electron and positron spectra and we will compute the angular distribution and the spectrum of the galactic synchrotron emission over a wide frequency range.  

Our approach aims at determining the interstellar spectrum of the CRE below 7 GeV under the condition that the Fermi-LAT, PAMELA and H.E.S.S.~lepton data are reproduced above that energy. We will therefore include the extra-component in our fits. We will pay particular attention also to the proton spectrum and to secondary-to-primary ratios by requiring that they are all reproduced in our models.
This will enable us to constrain relevant properties of the source spectrum and of the propagation conditions in the Galaxy.

The other main goal of this work is to constrain the vertical scale height of the diffusion region in the Galaxy. We will show that in addition to the traditional probe $^{10}{\rm Be}/ ^{9}$Be, other three observables can be exploited and can provide us with valuable insights. Two of them yield global (non-local) constraints. The three observables are:
\begin{itemize}
\item the radio spectrum, together with the condition that the GMF strength be compatible with Faraday rotation surveys;
\item the latitude profile of the synchrotron emission;
\item the positron spectrum at energies below $\sim 5~\GeV$.   
\end{itemize}

This paper is structured as follows. In section \ref{sec:GMF} we will discuss the characteristics of galactic magnetic fields. In section \ref{sec:CRspectra} we will study the spectra of CRE and of the galactic synchrotron emission. In section \ref{sec:b_profile} we will focus instead on the latitude profiles of the galactic synchrotron emission and on the problem of the scale height of the turbulent magnetic field $z_{t}$. Finally, in section \ref{sec:discussion} we will discuss our findings and draw our conclusions.

\section{The Galactic Magnetic Field}\label{sec:GMF}

A description of the \dragon\ code can be found in \cite{Evoli:2008dv} and on the \dragon\ webpage.~\footnote{\url{http://dragon.hepforge.org/}.} This code was successfully used to model CR nuclei and lepton propagation in the Galaxy and to compute the local interstellar spectra (LIS) of a number of secondary species including light nuclei, antiprotons, positrons and $\gamma$-rays. 

Differently from other semi-analytical and numerical codes, \dragon\ allows for a possible spatial dependence of the diffusion coefficient $D$ which we assume to take the general form:
\begin{equation}
\label{eq:diff_coeff}
D(\rho,R,z) = D_0 ~\beta^\eta f(z) \left(\frac{\rho}{\rho_0}\right)^\delta \;,
 \end{equation}
with $\rho \equiv p\beta c/(Ze)$ being the rigidity of the particle of charge $Z$ and momentum $p$, $f(z)$ describes the spatial dependence (in cylindrical coordinates and assuming azimuthal symmetry, with $z = 0$ on the Galactic plane and $f(0)=1$) of $D$, and $\eta$ provides an effective handling of the low energy behavior of $D$. 
While one would expect  $\eta=1$ as the most natural dependence of diffusion on the particle speed, several effects may give rise to a different effective behavior. Here, where not differently stated, we tune $\eta$, as other parameters, by minimizing the $\chi^2$ of the model against B/C and proton data (see below).

In previous versions, a simplified single component and $z$-independent model of the GMF was assumed to compute synchrotron energy losses of CR leptons, as that was enough to model them with sufficient accuracy. That approximation, however, is not sufficient when trying to get a realistic description of the synchrotron emission angular distribution as we want to do in this work. Therefore we adopt here a more realistic model of the GMF based on Faraday rotation measurements (RMs) of a large number of Galactic and extragalactic radio sources (see \cite{Han:2009ts} for a recent review on GMF). 

We consider here two main field components. For the regular component we adopt the recent model described in \cite{Pshirkov:2011um} and vary the main parameters in the allowed range (tables I and II in that paper). We notice that only the \emph{halo component} of the regular GMF has a role for our analysis, albeit marginal. This has a double torus structure, each of them with thickness $\sim 0.2-0.4~\kpc$ and at distance of $1 - 2~\kpc$ from the Galactic plane (GP).  
Its strength on the torus axis was estimated to be in the range $B_{\rm halo} \sim 2 - 12~\muG$ with $B_{\rm halo} \sim 4~\muG$ as a best fit.  
Recently, a rather different model of the regular component of the GMF has been proposed \cite{Jansson:2012pc}, which features, besides the spiral arms and disk and halo components, also an $X$-shaped component in the $r-z$ directions, especially relevant close to the galactic center. This GMF is explicitly divergence-free.~\footnote{The study of the random component associated with this model is in progress \cite{Glennys_private}.} We checked that, as expected, changing from one to the other model does not affect significantly our results. Therefore, we will use the model of \cite{Pshirkov:2011um} in the rest of this paper.
 
Very little is known on the geometrical structure of the random component of GMF.   
Its energy power spectrum has been determined on the basis of RMs to be \cite{Han:2004aa}
\begin{equation}
E_B(k) = C \left(\frac{k}{k_0}\right)^\alpha
\end{equation}
with $\alpha = -0.37 \pm 0.1$ and $C = (6.8 \pm 0.8) \times 10^{-13}~{\rm erg~ \cm^{-3}}$ in the wavenumber range $0.07 < k < 2~\kpc^{-1}$, corresponding to a rms field $B_{0} = 6.1 \pm 0.5~\muG$.
This measurement refers to the Galactic plane region where most of the pulsars are observed. At smaller scales the field power spectrum was found to be steeper, compatible with a Kolmogorov one. Its normalization matches the extrapolation down to $k \sim 1~\kpc$ of the spectrum measured at larger scales. 

Random GMFs are responsible for CR diffusion, hence the spatial and rigidity dependence of the diffusion coefficient should be related to the turbulent GMF spatial distribution and power spectrum, respectively. This is supported by quasi-linear theory and numerical simulations of particle propagation in turbulent MFs  (see e.g.~\cite{DeMarco:2007eh}).  

Concerning the dependence upon rigidity, we remark that it does not need to be the same as the one determined from the turbulence spectrum within the quasi-linear theory, as noticed in \cite{DeMarco:2007eh}: therefore, we leave the power-law index $\delta$, defined in eq.~\ref{eq:diff_coeff}, as a free parameter.

We assume instead the same $z$ dependence for $B_{\rm ran}$ and $D^{-1}$:  
\begin{equation}
D(z)^{-1} \propto B_{\rm ran}(z) \propto \exp\{- z/z_t\}~,
\label{eq:D_vs_z}
\end{equation}
where $z_{t}$ takes the role of the effective scale-height of the diffusion region (in works by other groups a constant diffusion coefficient is assumed, therefore the scale-height $z_{t}$ and the position of the vertical boundary of the propagation region $L$  coincide). In order to ensure that boundary effects are not relevant, we solve the transport equation in \dragon\ by setting always the vertical boundary $L=3z_{t}$.

\section{Cosmic-ray electron, positron and synchrotron spectra}\label{sec:CRspectra}

In this section we use \dragon\ to model the CR electron and positron spectra and to consistently compute the synchrotron emission of the Galaxy integrated along the line of sight. 

\begin{figure}[tbp]
\begin{center}
\includegraphics[width=0.47\textwidth]{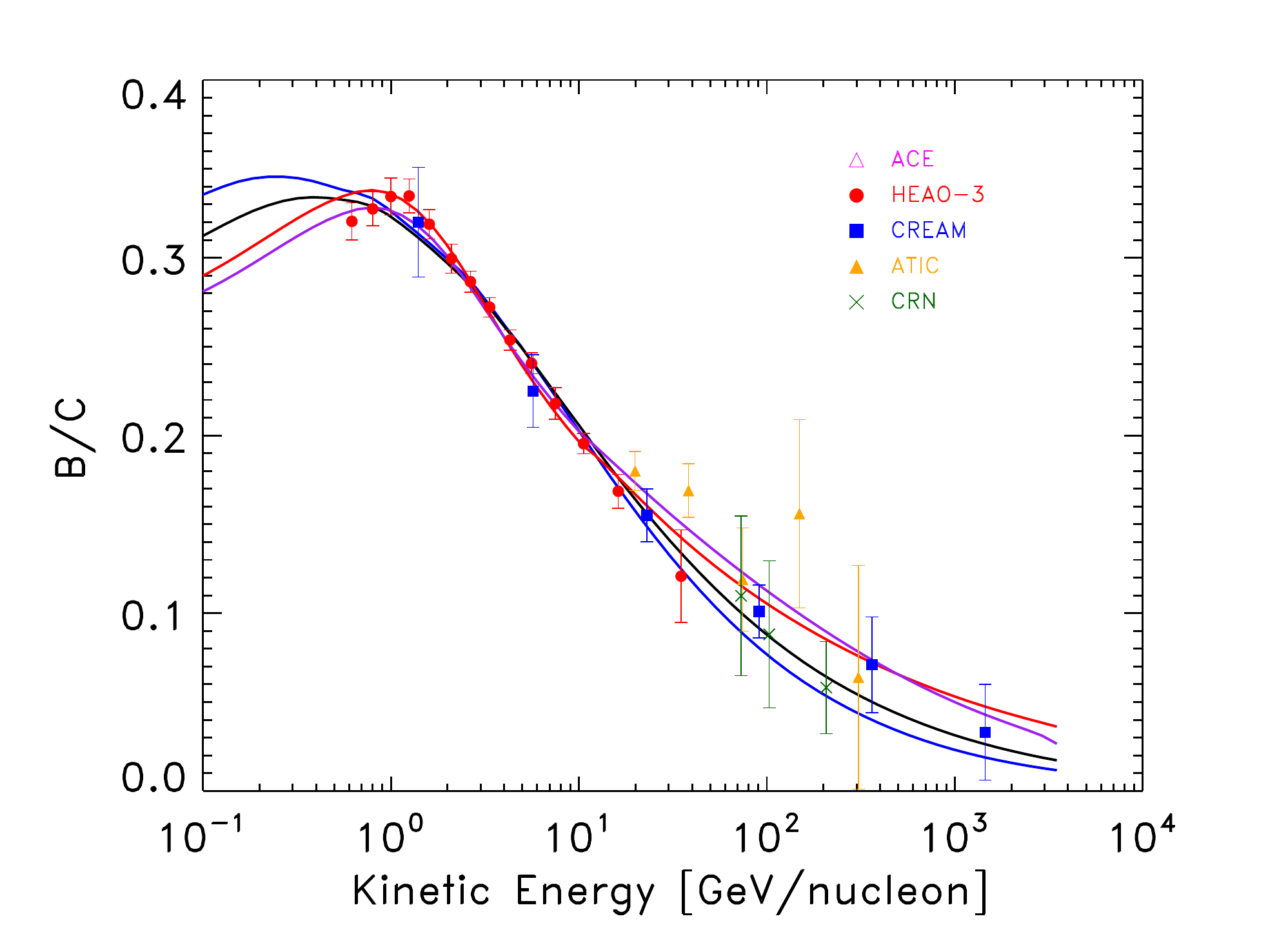}
\includegraphics[width=0.47\textwidth]{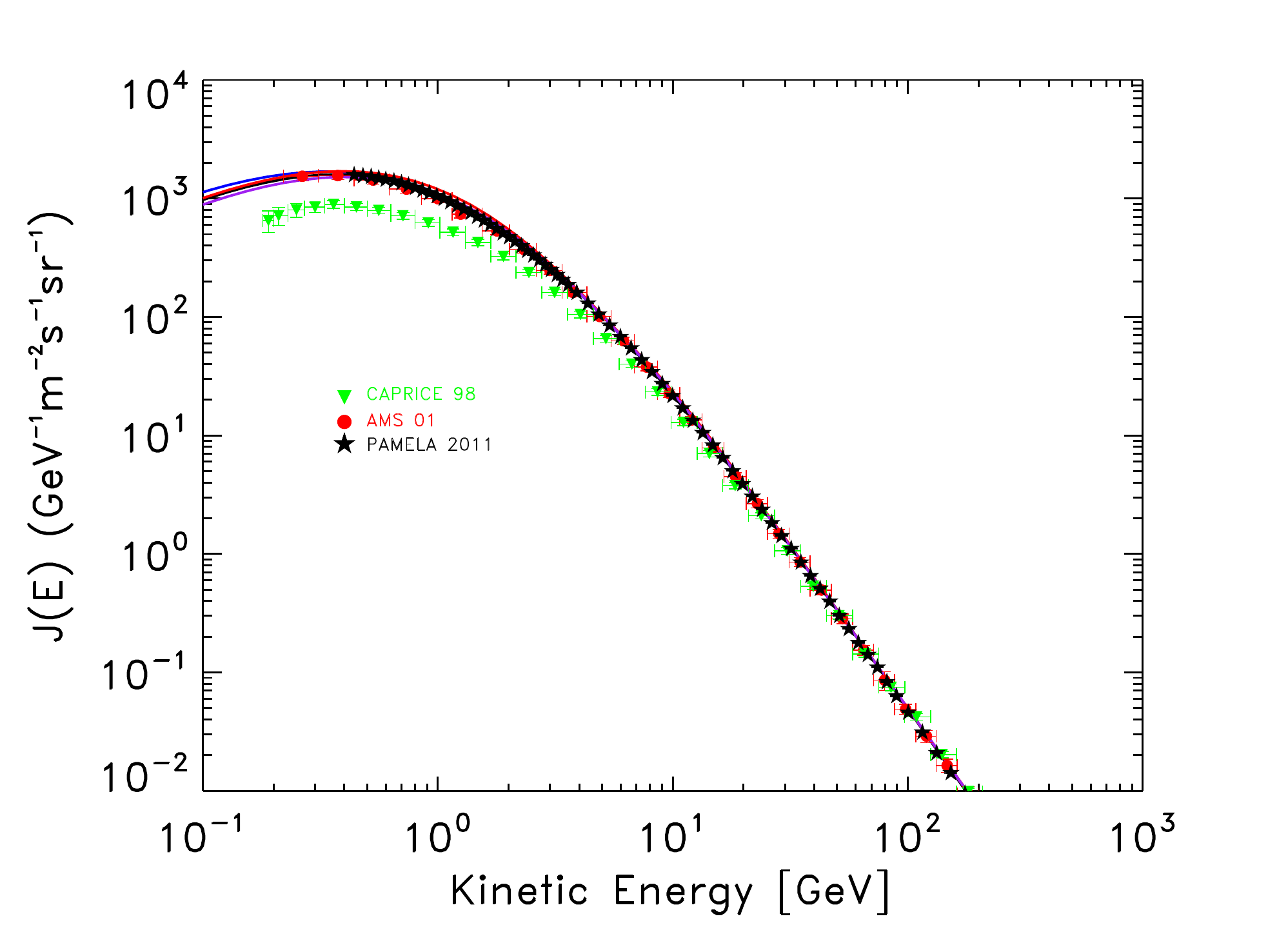}
\caption{B/C and proton spectra computed for the reference propagation setups PD4 (black lines),  KRA4 (blue) KOL4 (red) and CON4 (purple) defined in table \ref{tab:setups} are compared with a selection of experimental data. The modulation potential used for the B/C is $\Phi = 0.5$ GV.}
\label{fig:BCprspectra}
\end{center}
\end{figure}

In order to assess the uncertainties related to CR propagation, we consider four representative propagation regimes: PD (plain diffusion), KRA (Kraichnan), KOL (Kolmogorov) and CON (convective). 
For each of them we vary the scale-height of the diffusive halo in the range $z_t = 1\div16~\kpc$. 
The main parameters of these setups for $z_{t}\leq8~\kpc$ are summarized in table \ref{tab:setups}.  
All these models are chosen in such a way as to minimize the combined $\chi^2$ against the B/C and the proton observed spectra, as shown in figure \ref{fig:BCprspectra}.  Their reduced  $\chi^2$ is always smaller than unity. This approach is similar to the one adopted in \cite{Evoli:2011id}, to which we refer the reader for more details including the used data set.    

Once the model parameters are fixed according to this procedure, we have still the freedom to choose the spectral index and normalization of the injection spectrum of the primary electrons and of the extra-component. We fix these parameters by fitting the $e^- + e^+$ spectrum and the positron fraction measured by Fermi-LAT. In contrast, the spectrum and normalization of the secondary electron and positron spectra are determined once the proton and helium spectra (and the propagation parameters) are fixed. 

Solar modulation is one of the major issues in modeling CR spectra below $\sim 20~\GeV$. 
The most common treatment assumes the \emph{force-field} charge independent approximation \cite{Gleeson_1968ApJ}. In this approach the effects of propagation in the solar system on the observed spectra of all electrically charged CR species are described in terms of the single, time-dependent parameter $\Phi$ (the so called \emph{modulation potential}), to be fixed against CR data. 
However, charge dependent drifts in the complex magnetic structure of the heliosphere may lead to significant deviations from this simple scenario (see e.g.~\cite{2011ApJ...735...83S,2012Ap&SS.339..223S} for recent discussions of this problem). 

In order for our analysis to be as independent of solar effects as possible, we tune the CRE source parameters by comparing the force field modulated $e^- + e^+$ spectra with Fermi-LAT data, which have been measured above 7 GeV. According to \cite{1996ApJ...464..507C}, and taking into account that data were taken during a quiet Sun phase, drifts effects are expected to be small. Below that energy we will model the $e^- + e^+$ LIS on the basis of the observed synchrotron spectrum of the Galaxy which is unaffected by propagation in the heliosphere. At this point the low energy CR electron and positron spectra at Earth will be \emph{predicted} (assuming a model for solar effects) and low energy data can be used to further constrain the model.

A comment is in order at this point: We could have chosen to consider the electron spectrum and the positron fraction measured by PAMELA in order to fix the relevant CRE parameters, and then compare with the synchrotron spectrum and the Fermi CRE spectra. We did not follow this route because: a) PAMELA data on the CRE spectrum do not extend to high enough energy to probe accurately the extra-component; b) PAMELA data extend instead down to low energies, where solar modulation effects can be significant. We remark that using a combination of the spectra observed by both experiments is not feasible either, because of the presence of a systematic offset between PAMELA and Fermi data, which would make extremely difficult a proper comparison between them. 

\begin{table}[tbp]
\centering
  \begin{tabular}{|c|c|c|c|c|c|c|c|c|c|c|}
    \hline
     \multirow{2}{*} {\bf Model} & $D_0$ & $v_A$  & $dV_c/dz$  & $\delta$ &  $\eta$ &  $\gamma$ & $\Phi_{p}$  \\
     &  $(10^{28}\cm^{2}/\s)$ & $(\km/\s)$ & $(\km/\s/\kpc)$ & & & & $(\rm GV)$ \\
    \hline
\hline    
    	PD8  &   3.64 & 0* & 0* & 0.56 & -0.42 & 2.3 &  0.57    \\
	PD4  &   2.24 & 0* & 0* & 0.57 & -0.40 & 2.29 & 0.56  \\
	PD2  &   1.11 & 0* & 0* & 0.58 & -0.45 & 2.28 & 0.57  \\
	PD1  &   0.515  & 0* & 0* &  0.58 &  -0.40 & 2.29 & 0.58  \\
\hline 
	KRA8  &  4.28 & 15.7 & 0* & 0.5* & -0.39 & 2.35 & 0.68   \\
	KRA4  &  2.60 & 13.7 & 0* & 0.5* & -0.39 & 2.36 & 0.65   \\
	KRA2  &  1.35 & 15.6 & 0* & 0.5* & -0.41 & 2.35 & 0.67  \\
	KRA1  &  0.627 & 15.3 & 0* & 0.5* & -0.40 &  2.33 & 0.69  \\
\hline
	KOL8 & 7.17 & 38.9 & 0* & 0.33* & 1* & 2.00/2.40  & 0.52    \\
	KOL4 & 3.98 & 33.4 & 0* & 0.33* & 1* & 1.93/2.47  & 0.51   \\		
	KOL2 & 2.07 & 31.9 & 0* & 0.33* & 1* & 1.91/2.47  & 0.54   \\
	KOL1 & 1.06 & 35.9 & 0* & 0.33* & 1* & 1.81/2.40  & 0.47    \\
	\hline
        CON8 & 0.952 & 36.1 & 50* & 0.58 & 1* & 1.92/2.48  & 0.52    \\      
        CON4 & 0.923 & 36.0 & 50* & 0.53 & 1* & 1.94/2.48  & 0.52     \\
        CON2 & 0.794 & 34.5 & 50* & 0.48 & 1* & 1.95/2.46  & 0.52    \\
	CON1 & 0.573 & 31.4 & 50* & 0.41 & 1* & 1.95/2.46  & 0.52  \\
     	\hline
  \end{tabular}
\caption{\label{tab:setups} We report here the main parameters of the reference propagation setups used in this work. The number appearing in the name of the model indicates the used value of $z_{t}$. The convective velocity $V_c$ is assumed to vanish on the Galactic plane $(z = 0)$. The KOL and CON models have a break in  the nuclei source spectra $\gamma$ at a rigidity of 11 GV. Parameters marked with a $*$ are fixed a priori to characterize the setup. More details on these models are reported on the \dragon\ webpage.}
\end{table}

After having computed with \dragon\ the CR electron and positron spatial distributions and energy spectra in the Galaxy, we use the standard formalism to compute the synchrotron emissivity at each point. For a monochromatic and isotropic distribution of electrons, this reads \cite{Longair,1988ApJ...334L...5G}
\begin{eqnarray}
\epsilon(\nu,\gamma)_{\rm reg} &=& \sqrt{3} \frac{e^{3}}{mc^{2}}B_{\rm perp}F(x) \label{eq:emissivity} \\
\epsilon(\nu,\gamma)_{\rm rand} &=& C y^{2}\left(K_{4/3}(y)K_{1/3}(y)-\frac{3}{5}y\left(K_{4/3}^{2}(y)-K_{1/3}^{2}(y)\right)\right)\;.
\end{eqnarray}
In eq.~\ref{eq:emissivity}, $\epsilon(\nu,\gamma)_{\rm reg}$ refers to the unpolarized emissivity at the frequency $\nu$ due to the regular magnetic field component perpendicular to the line of sight $B_{\rm perp}$, from electrons with energy $E=mc^{2}\gamma$. We define $x=\nu/\nu_{c}^{\rm reg}$, where $\nu_{c}^{\rm reg}=3e/(4\pi mc)B_{\rm perp}\gamma^{2}$ is the critical synchrotron frequency, while $F(x) = x\int_{x}^{\infty}K_{5/3}(x')dx'$. The functions $K_{l}$ are modified Bessel functions or order $l$. In the expression of the isotropic emissivity due to the random component we have $C=2\sqrt{3}e^{3}/(mc^{2})B_{\rm rand}$, where $y=\nu/\nu_{c}^{\rm rand}$ and $\nu_{c}^{\rm rand} = 3e/(2\pi mc)B_{\rm rand}\gamma^{2}$. For each line of sight we then integrate the emissivity to obtain the intensity $I(\nu)$.

Sometimes, we will refer to the brightness temperature, which is defined as
\begin{equation}
T = \frac{I(\nu)c^{2}}{2\nu^{2}k_{B}}\;,
\end{equation}
where $k_{B}$ is the Boltzmann's constant.

In the rest of this section we will keep the diffusive halo scale-height fixed at the reference value $z_t = 4~\kpc$.
The effect of varying this parameter will be discussed in section \ref{sec:b_profile}.

\subsection{CRE models based on Fermi-LAT and synchrotron spectrum data}\label{sec:FERMI_fit}

As we mentioned in the Introduction (see also \cite{Grasso:2009ma,Ackermann:2010ij,Grasso:2011wt}), an acceptable fit can be found only if, besides a standard electron $e^-$ component with a power-law source spectrum $J_0^{e^-} \propto E^{-\gamma_0(e^-)}$, we introduce an electron and positron charge symmetric extra-component with the form $J_0^{e^{\pm}}(E) \propto E^{-\gamma(e^\pm)} \exp \left(-E/E_{\rm cut}\right)$ with $E_{\rm cut} \simeq 1~\TeV$.  
We assume that the spatial distribution of both the standard and extra-component sources traces that of SNRs as determined from pulsar catalogues. 
In section \ref{sec:discussion} we will briefly discuss how our results change assuming the extra-component to be local hence not contributing to the large scale Galactic synchrotron emission. 
We also account for secondary electrons and positrons produced by CR nuclei scattering onto the ISM (only $p-p$, $p-{\rm ^4He}$ and $^4{\rm He}-{\rm ^4He}$ are relevant) adopting the parameterization of the production cross sections provided in \cite{Kamae:2006bf}. The cross sections provided in \cite{Huang:2007wk} do not differ from the ones in \cite{Kamae:2006bf} in the energy range relevant to our work \cite{Pohl_private}.

\begin{figure}[tbp]
\centering
\includegraphics[width=0.6\textwidth]{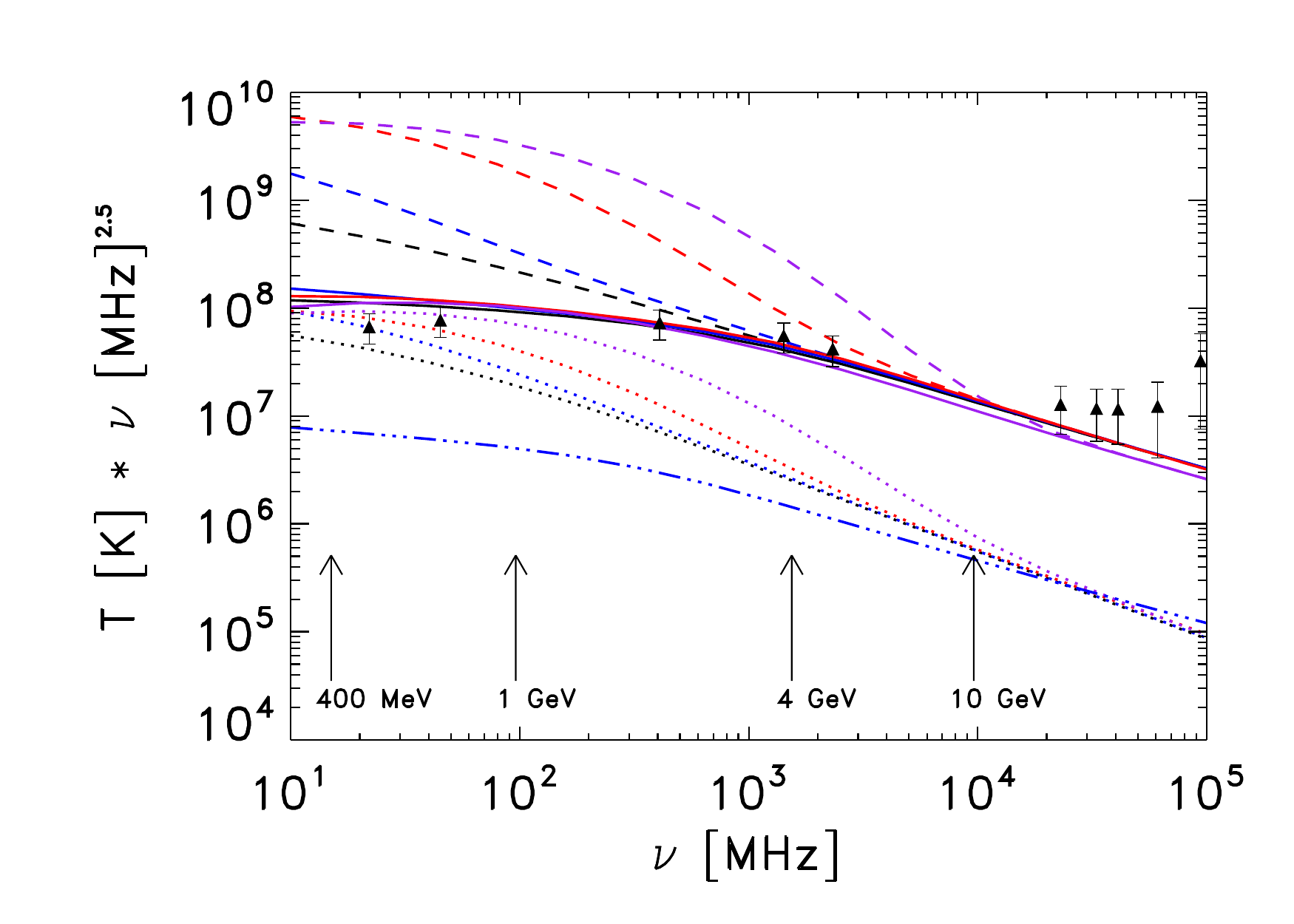}
\caption{The average synchrotron spectra in the region $40^\circ < l < 340^\circ$, $10^\circ < b < 45^\circ$ computed for the reference propagation setups PD4 (black lines),  KRA4 (blue) KOL4 (red) and CON4 (purple) defined in table \ref{tab:setups} are compared with experimental data derived from \cite{Oliveira-Costa_2008MNRAS}. For each setup we show the spectra obtained with (solid lines) and without (dashed line) the spectral break in the $e^-$ source spectra given in table \ref{tab:CRE_models}. Dotted lines represent the corresponding contribution of secondary $e^-$ and $e^+$. The contribution to the synchrotron flux of the regular GMF with $B_{\rm halo} = 4~\muG$ (see \cite{Pshirkov:2011um}), computed for the KRA4 setup, is shown as the triple dot-dashed line. The random component field stength is tuned to reproduce the spectrum normalization at 408 MHz. The required normalization is $B_{\rm ran}(0) = 7.6~\muG$. The critical synchrotron frequencies, calculated for this value of $B_{\rm ran}(0)$, are reported for a few reference values of the electron energy.  
 Microwave data above 20 GHz are expected to be contaminated by non-synchrotron emission. Therefore they only provide upper limits to the synchrotron flux and are shown here only as a reference.}
\label{fig:synchro_vs_setups} 
\end{figure}

We then compute the synchrotron spectrum of the Galaxy for the CRE models obtained as described above. For each of these models we tune consistently, through an iterative procedure, the normalization of the random component of the GMF in order to fit the observed synchrotron spectrum at 408 MHz where radio data are most complete and a possible contamination due to {\it free-free} emission is negligible \cite{Oliveira-Costa_2008MNRAS}.  
For the regular component of the GMF we adopt model \cite{Pshirkov:2011um} (see section \ref{sec:GMF}) which was built on Faraday rotation measurements. 
In figure \ref{fig:synchro_vs_setups} we show for the KRA4 that the contribution of that field component to the synchrotron spectrum is subdominant. The same holds for the other setups. As a consequence, the effect of changing the regular GMF model among those discussed in section \ref{sec:GMF} is negligible as we checked also for different values of $z_t$.
 
We integrate the Galactic emission along the line of sight and use {\tt HEALPix}~\footnote{\url{http://healpix.jpl.nasa.gov}.} to properly average the resulting flux over the sky regions $40^\circ < l < 340^\circ$, $10^\circ < b < 45^\circ$ and $-45^\circ < b < -10^\circ$ where $l$ and $b$ are Galactic longitude and latitude respectively. This is the region where the contamination from point-like and local sources is expected to be the smallest and it is the same considered in \cite{Strong:2011wd}. 
In this region we compare the simulated spectra with the ones measured by a wide set of radio surveys  at 22, 45, 408, 1420 , 2326 MHz 
as well as WMAP foregrounds at 23, 33, 41, 61 and 94 GHz as consistently catalogued in \cite{Oliveira-Costa_2008MNRAS} (to which we refer the reader for the survey references) where they were pixelized and transformed in Galactic coordinates.  The Cosmic Microwave background is subtracted from WMAP data. 
Above 1 GHz the \emph{free-free} emission was estimated to give a non-vanishing contribution; 
the 94 GHz channel is likely to be dominated by thermal dust-emission. Therefore data above a few GHz provide only {\em upper limits} to the flux of synchrotron radiation reaching the Earth.  
 Those high frequency data, however, will not be used in our analysis since the CRE spectrum above 7 GeV (corresponding to frequencies larger than a few GHz)  will be modeled only on the basis of Fermi-LAT and H.E.S.S. data.  
Absorption due to \emph{free-free} scattering is estimated to be negligible down to $10$ MHz \cite{Rybicki}. 
The error bars on the radio fluxes reported in our figures correspond to the flux semi-dispersion in the considered sky region. 
We remove the Galactic plane, as well as discrete and local diffuse radio sources (e.g.~the Galactic Spur) using the extended temperature analysis mask used by the WMAP collaboration.~\footnote{\url{http://lambda.gsfc.nasa.gov/product/map/dr4/masks\_get.cfm}.} The CRE models obtained according to this procedure are shown in table \ref{tab:CRE_models}. 

\begin{table}[tbp]
\centering
  \begin{tabular}{|c|c|c|c|c|c|c|}
    \hline
    {\bf Model} &  $\gamma_0(e^-)$ &  $\gamma(e^\pm)$  & $E_{\rm cut}~(\GeV)$ & $\Phi~(\rm GV)$ & $\chi^{2}_{\rm FERMI}$  \\
    \hline
     PD4 & 0.9/2.53 &  1.38 &  986 & 0.3 & 0.128 \\
     KRA4 & 0.6/2.53  & 1.50 & 967 & 0.3 & 0.20 \\
     KOL4 &  0.5/2.54  & 1.56 & 985 & 0.1& 0.177  \\
     CON4 &  0.5/2.45 & 1.45 & 999 & 0.1& 0.274  \\
    \hline
  \end{tabular}
\caption{ \label{tab:CRE_models} We report here the source parameters which yield a consistent fit of the CRE and synchrotron spectra. The reported values of low energy spectral indexes are the maximum allowed to reproduce radio data at $2\sigma$. The values of the force field modulation potential $\Phi$ have been fixed to match the $e^-$ AMS-01 data.  The reduced $\chi^2$  is computed against the $e^+ + e^-$ spectrum and the positron fraction measured by Fermi-LAT.}
\end{table}

Figure \ref{fig:synchro_vs_setups} (solid lines) shows a set of representative models which fit Fermi-LAT data and provide simultaneously a good description of the measured synchrotron spectrum. 
 Noticeably, for all setups with $z_t = 4~\kpc$ the required random field strength $B_{\rm ran}(0) = 7.6~\muG$ for $10^\circ < b < 45^\circ$ and $6.6~\muG$  for $-45^\circ < b < - 10^\circ$. In both cases the field strength is well within the interval allowed by RM (see section \ref{sec:b_profile} for a wider discussion on this issue). 

We notice, however, that the experimental data are reproduced only if a sharp spectral hardening, a break, is imposed to the $e^-$ source spectrum below $E \simeq 4~\GeV$. Its amplitude must be $\Delta \gamma(e^-) \simeq 1.6 (1.9) $ for the PD (KRA) setups and $\Delta \gamma(e^-) \simeq 2.0$ for the KOL and CON (see table \ref{tab:CRE_models}). This is similar to what was found in \cite{Strong:2011wd} with $\Delta \gamma(e^-) =  0.9$,  and in \cite{Bringmann:2011py} where $\Delta \gamma(e^-) = 1.57$.  We find some significant differences with respect to these papers though.

In contrast to \cite{Bringmann:2011py} we find the required break to be significantly dependent on the propagation setup (see table \ref{tab:CRE_models}). However, in \cite{Bringmann:2011py} the break of the \emph{propagated} spectrum was quoted, rather than that of the \emph{injected} spectrum. It is quite natural that the way in which the break translates into the other depends on the propagation model. Moreover, the break quoted in \cite{Bringmann:2011py} is that of the \emph{total} (primary+secondary) propagated spectrum. However, the contribution of secondary electrons and positrons depends on the primary proton and helium spectra in addition to the propagation setup (see the dotted lines in figures \ref{fig:synchro_vs_setups} and \ref{fig:models_FERMI}), and cannot be arbitrarily altered.  Similarly to what we noticed in \cite{DiBernardo:2010is} (see also \cite{1998ApJ...493..694M} for an analogous remark), the spectra of secondary particles display a bump at $E \sim 1~\GeV$. The bump is particularly prominent for models with strong reacceleration, as the KRA and the CON setups. 
Due to the presence of this component, the observed synchrotron and CRE spectra can be reproduced only by imposing a strong break or a cutoff in the primary $e^-$ component.

Differently from \cite{Strong:2011wd} we find a larger flux of secondary particles in the GeV range and that a more pronounced break is required.
We notice that the values of the low-energy spectral indexes reported in table \ref{tab:CRE_models} and adopted in figure \ref{fig:synchro_vs_setups} (solid lines), are \emph{maximal} values, i.e.~they correspond to the minimal breaks, allowed by the radio data at $2\sigma$.
Indeed we find that assuming lower values of $\gamma_0(e^-)$, or replacing the break with an exponential infrared cutoff of the primary $e^-$ component at $E_{\rm low} \sim 2 ~\GeV$, do not change our results significantly.  
Therefore, below $\sim 100$ MHz the synchrotron spectrum of the Galaxy becomes gradually dominated by secondary $e^-$ and $e^+$. 

\begin{figure}[tbp]
\centering
\includegraphics[width=0.6\textwidth]{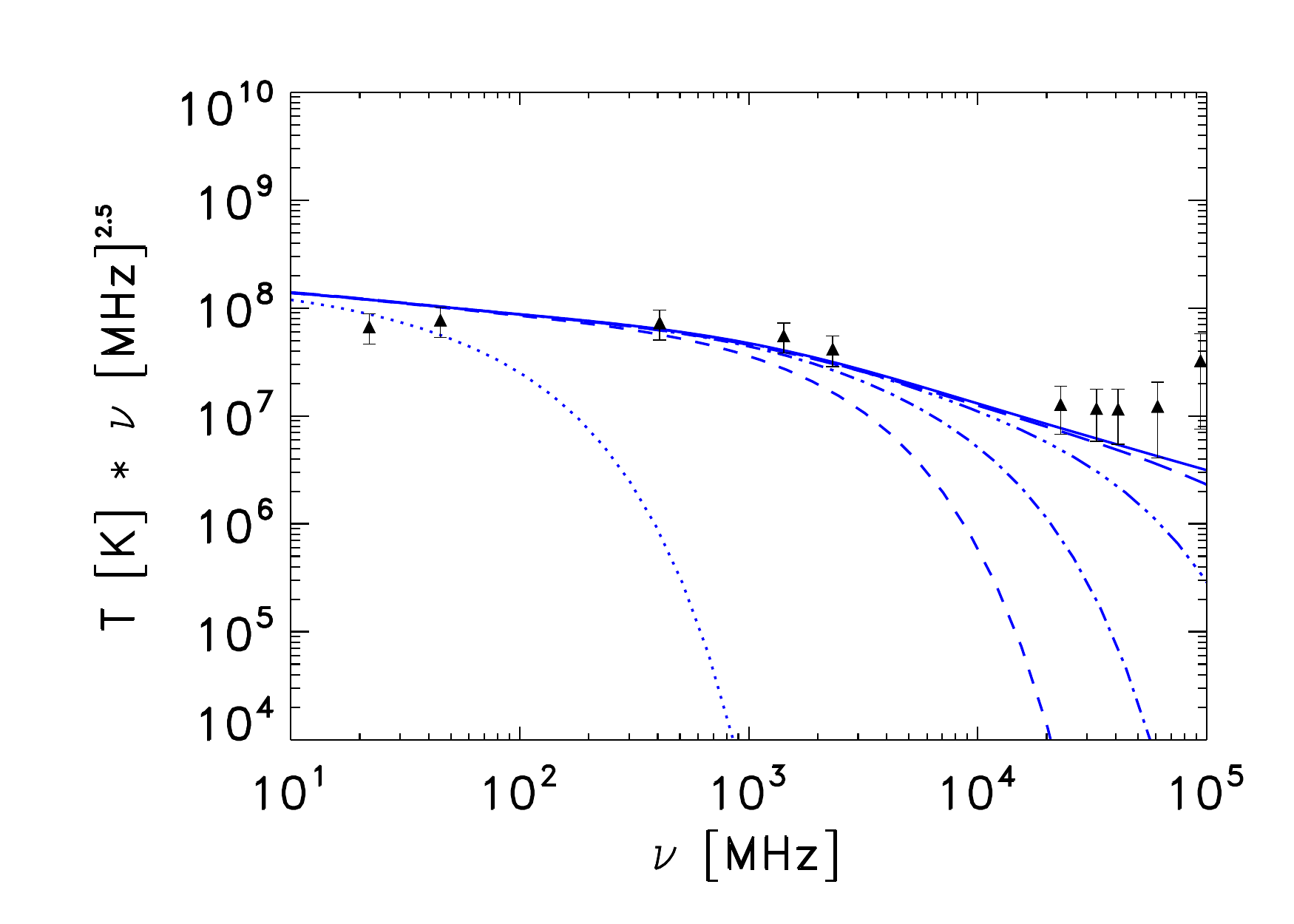}
\caption{The contribution to the synchrotron spectrum of CRE with energies $E < 1, 5, 10, 20, 50, 1000~\GeV$ (from the bottom to the top) is shown for the KRA4 model. The curves are computed in the same sky region and for the same magnetic field used in Fig. \ref{fig:synchro_vs_setups}}. 
\label{fig:Ecut_spectrum} 
\end{figure}

 In order to clarify which CRE energies are most relevant in the considered frequency range, in Fig.~\ref{fig:Ecut_spectrum} we show the synchrotron spectrum generated by CRE with energies $E < 1,5,10,20, 1000~\GeV$. It is evident that the electron + positron extra-component, which dominates the CRE spectrum above a few hundred GeV, gives a negligible contribution in the frequency range considered in this work. 

\subsection{\boldmath Back to the $e^+$ and $e^-$ observed spectra}\label{subsec:cre_spectra}

In the last part of this section we turn our attention back to the electron and positron spectra measured in the solar system. Our aim is to check to which extent the models we built to describe the synchrotron emission are consistent with CRE data even below 7 GeV. This results will provide a valuable consistency check for the validity of our approach and of the force field modulation treatment that we use in this paper.

We only consider 
data taken only during periods of low solar activity since under those conditions force field modulation is expected to provide the most reliable results. 
Our reference data sets below $7~\GeV$ are those taken by AMS-01 \cite{Aguilar:2007yf} in June 1998 and PAMELA between July 2006 and January 2010, both during low activity periods. We notice that the latter period is different from that during which PAMELA measured the proton spectrum which we used to tune our propagation setups (2006-08). This explains why the modulation potentials reported in tables \ref{tab:setups} and \ref{tab:CRE_models} are different. 

\begin{figure}[tbp]
\centering
\subfigure[] {
\includegraphics[width=0.47\textwidth]{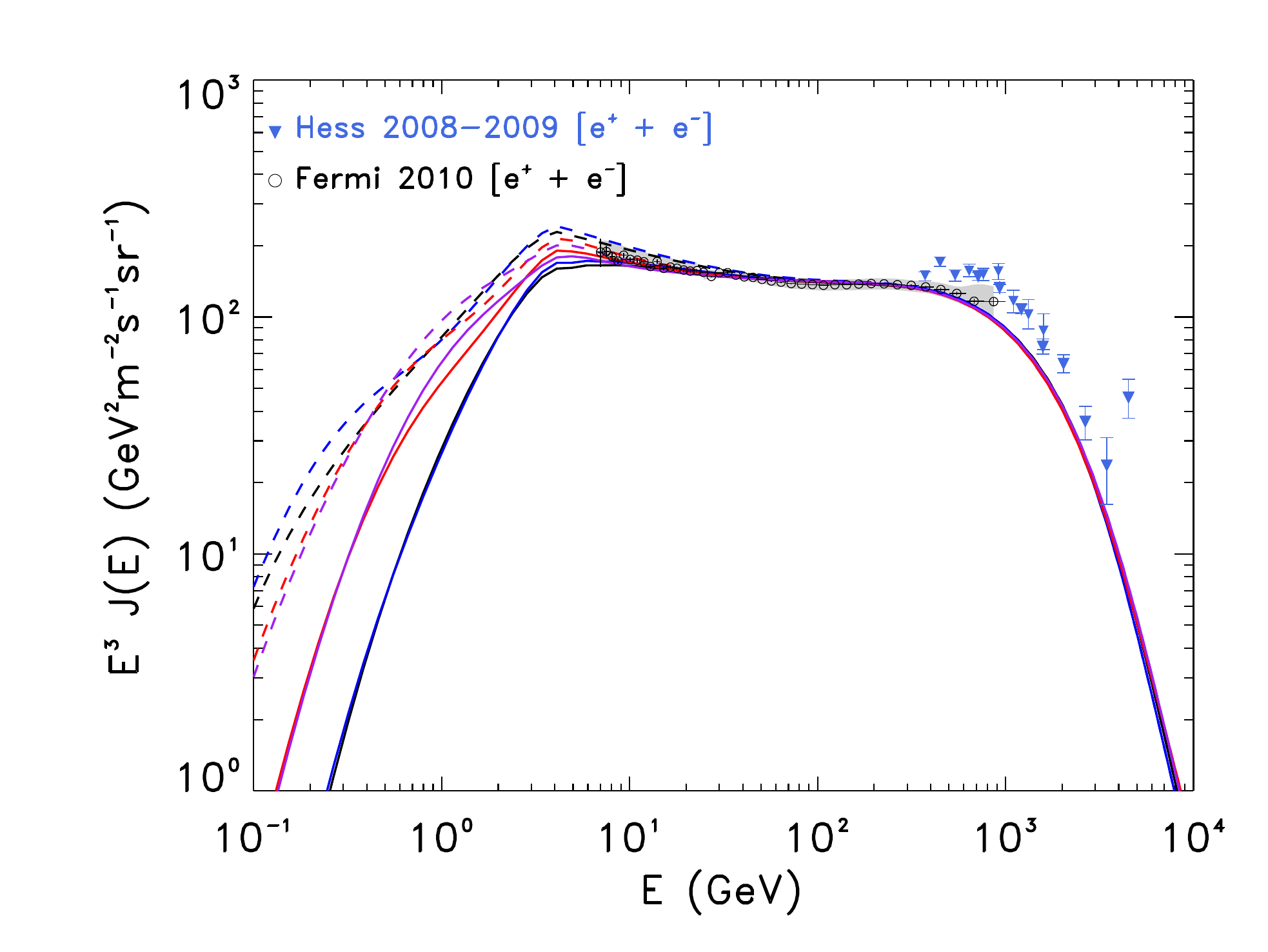}
}
\subfigure[] {
\includegraphics[width=0.47\textwidth]{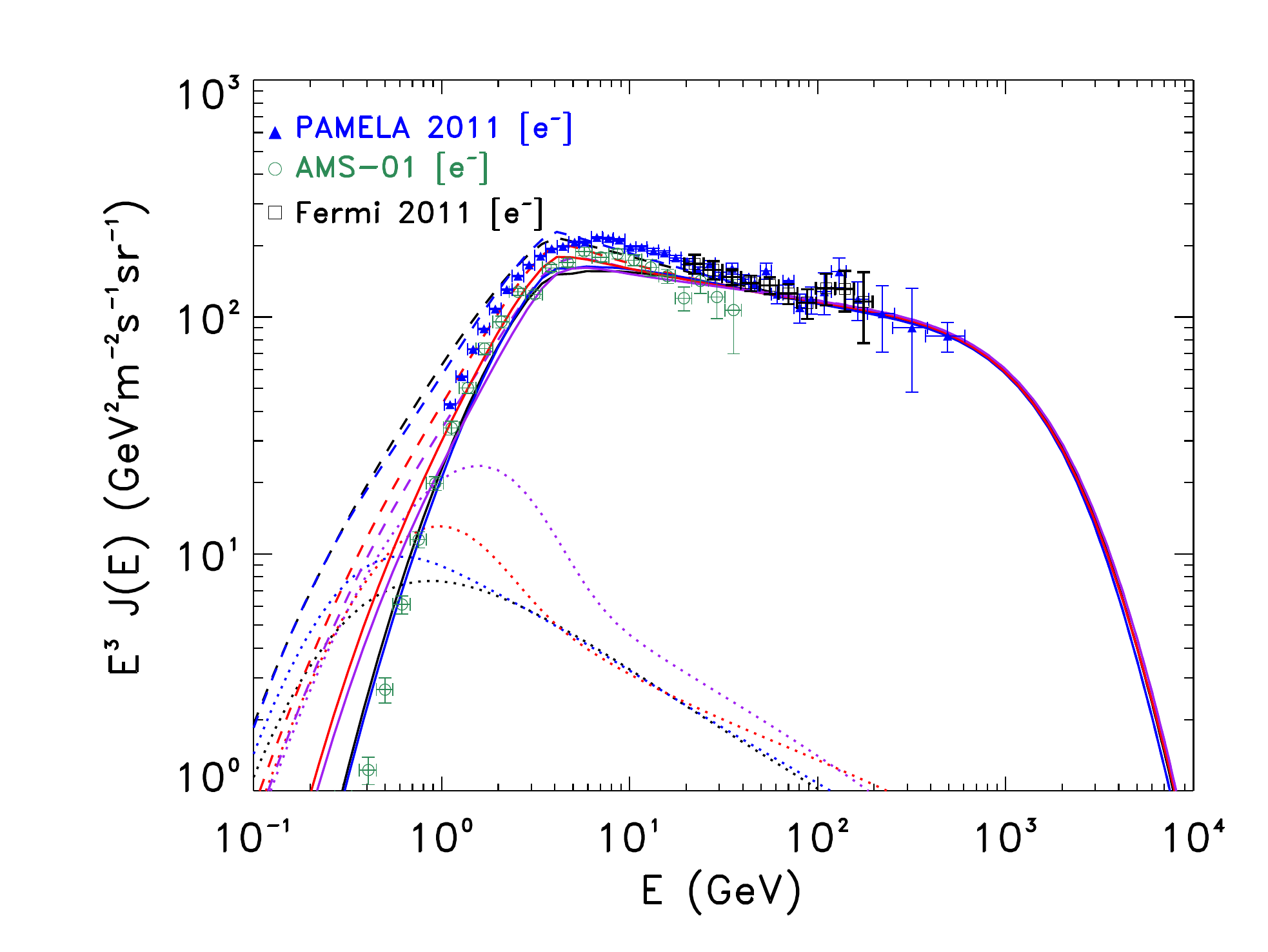}
}
\subfigure[] {
\includegraphics[width=0.47\textwidth]{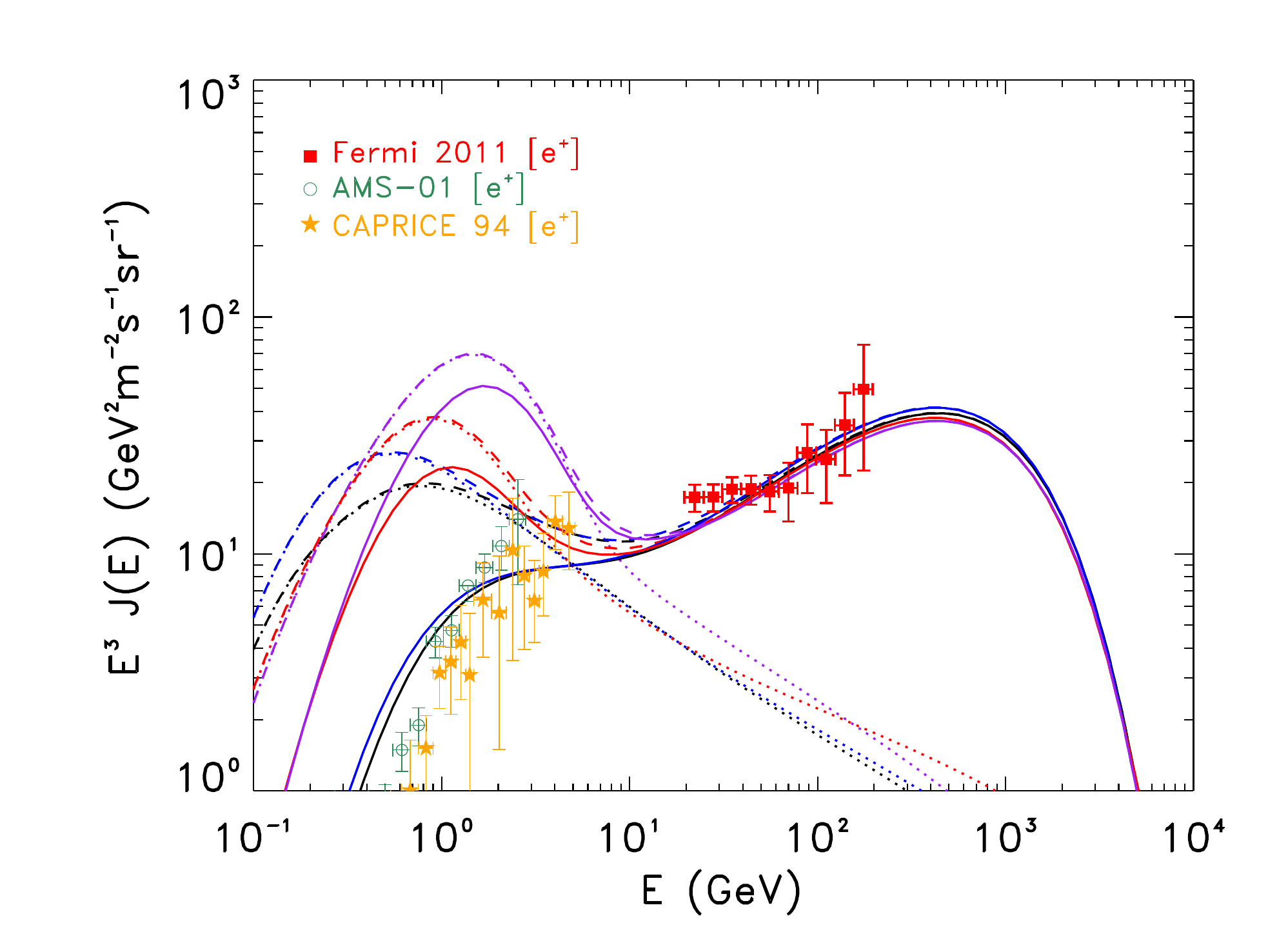}
}
\subfigure[] {
\includegraphics[width=0.47\textwidth]{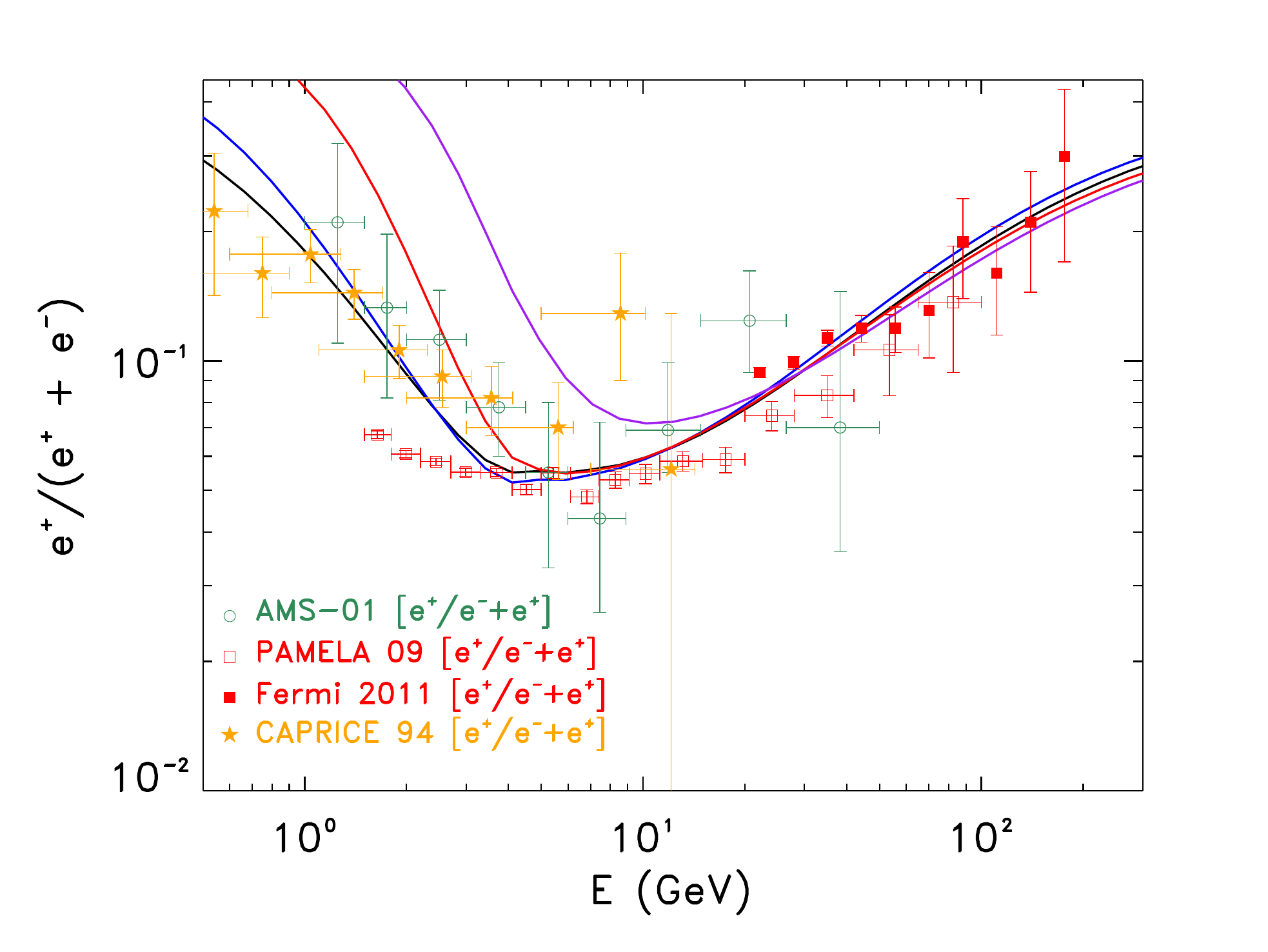}
}
\caption{The $e^+ + e^-$ (panel $a$), $e^-$ (panel $b$) and $e^+$ (panel $c$) spectra multiplied by $E^3$ as well as the positron fraction (panel $d$) 
computed for the reference propagation setup PD4 (black lines), KRA4 (blue), KOL4 (red) and CON4 (purple) defined by the parameters in tables \ref{tab:setups} and \ref{tab:CRE_models} are shown together with a selection of experimental data sets. 
Continuos and dashed lines represent modulated and interstellar (LIS) spectra respectively. The dotted lines correspond to the secondary contributions to the LIS spectra.}
\label{fig:models_FERMI}
\end{figure}

For each propagation setup, Fermi-LAT data can be fitted for several combinations of the $e^-$ source spectral index and the modulation potential $\Phi$. Low energy electron and positron data, however, break this degeneracy. The reason is that, as follows from our previous considerations, in the GeV and sub-GeV regions the $e^-$ and $e^+$ fluxes receive large, or dominant, contributions from secondary particles which are unaffected by the primary source spectrum and, in addition, the total spectrum is fixed by the synchrotron spectrum. The modulation potential is then fixed against AMS-01 data. In conclusion, for each setup the models in table \ref{tab:CRE_models} are unambiguously fixed by the requirement to match the observed CRE and synchrotron spectra.
Figure \ref{fig:models_FERMI} show that all models nicely reproduce the $e^-$ spectrum measured by AMS-01 down to $\sim 1~\GeV$. PAMELA $e^-$ data are also reproduced if we allow for a normalization offset compatible with the systematic errors of PAMELA and Fermi-LAT (relative to which our models are normalized).
 
Our findings for the positron flux are even more interesting. The $e^+$ spectrum below $10~\GeV$ strongly depends on the propagation setup. The KOL and CON configurations are clearly incompatible with AMS-01 data because of the presence of a pronounced bump at $E \sim 1~\GeV$, arising due to the strong reacceleration (see also \cite{DiBernardo:2010is}). The PD4 and KRA4 models, on the contrary, match the data more closely. 

For what concerns the positron fraction, there is a clear (and well-known) discrepancy between AMS-01 and PAMELA data below a few GeV. While this may be partially due to a normalization offset between the two experiments, 
the different slopes at low energies can only be explained by solar modulation and the drift of the particles in the heliosphere \cite{DellaTorre:2012zz}.
Indeed, these experiments were operating during periods with opposite polarity of the heliospheric magnetic field. 
It is a quite generic feature of drifts in the solar system to either keep particles diffusing tight to the solar current sheet (when the charge sign and the polarity are opposite) or to push them outside the current sheet (when the charge and the polarity have the same sign) \cite{2011ApJ...735...83S,2012Ap&SS.339..223S}. This will therefore increase or decrease the propagation time in the heliosphere, hence enhancing or lowering the energy losses, respectively. 

When dealing with ratios like the positron fraction, the situation is even more complicated, because the effects described above mix (an account of these arguments related to the very similar case of the antiproton/proton ratio can be found in \cite{0004-637X-565-1-280} and references therein). 
Interestingly, it was found that in phases of low solar activity and positive polarity, as that when AMS-01 was operating, the positron fraction computed with the force field approximation and a small modulation potential agrees quite well with those models \cite{DellaTorre:2012zz}. 
Our models are based on high energy CRE data and synchrotron spectra, hence they are not affected by drift effects in the solar system. Therefore, the agreement with AMS-01 data down to $\sim 1~\GeV$ provides a valuable check of the consistency of our approach. 

We notice that the KRA model  in \cite{DiBernardo:2009ku}, though similar to its homologous in this paper, reproduced the positron fraction measured by PAMELA down to $1~\GeV$ rather than that of AMS-01. That model, however, adopted a weaker break of the $e^-$ source spectrum as the synchrotron emission of the Galaxy was not fitted there.  This shows the relevance of the latter channel to understand low energy CRE spectra. 

We conclude that low reacceleration models with a strong flattening, or a IR cutoff, in the $e^-$ injection spectrum below a few GeV are consistent with the full set of radio and CR direct measurements.  

\section{Halo height and the latitude profile of the synchrotron emission}\label{sec:b_profile} 

\subsection{\boldmath The effects of varying $z_t$ on the synchrotron spectrum and angular distribution}

We study here how our previous results depend on the scale height $z_t$ of the GMF random component and place constraints on this quantity on the basis of available experimental data. 

\begin{figure}[tbp]
\centering
\includegraphics[width=0.47\textwidth]{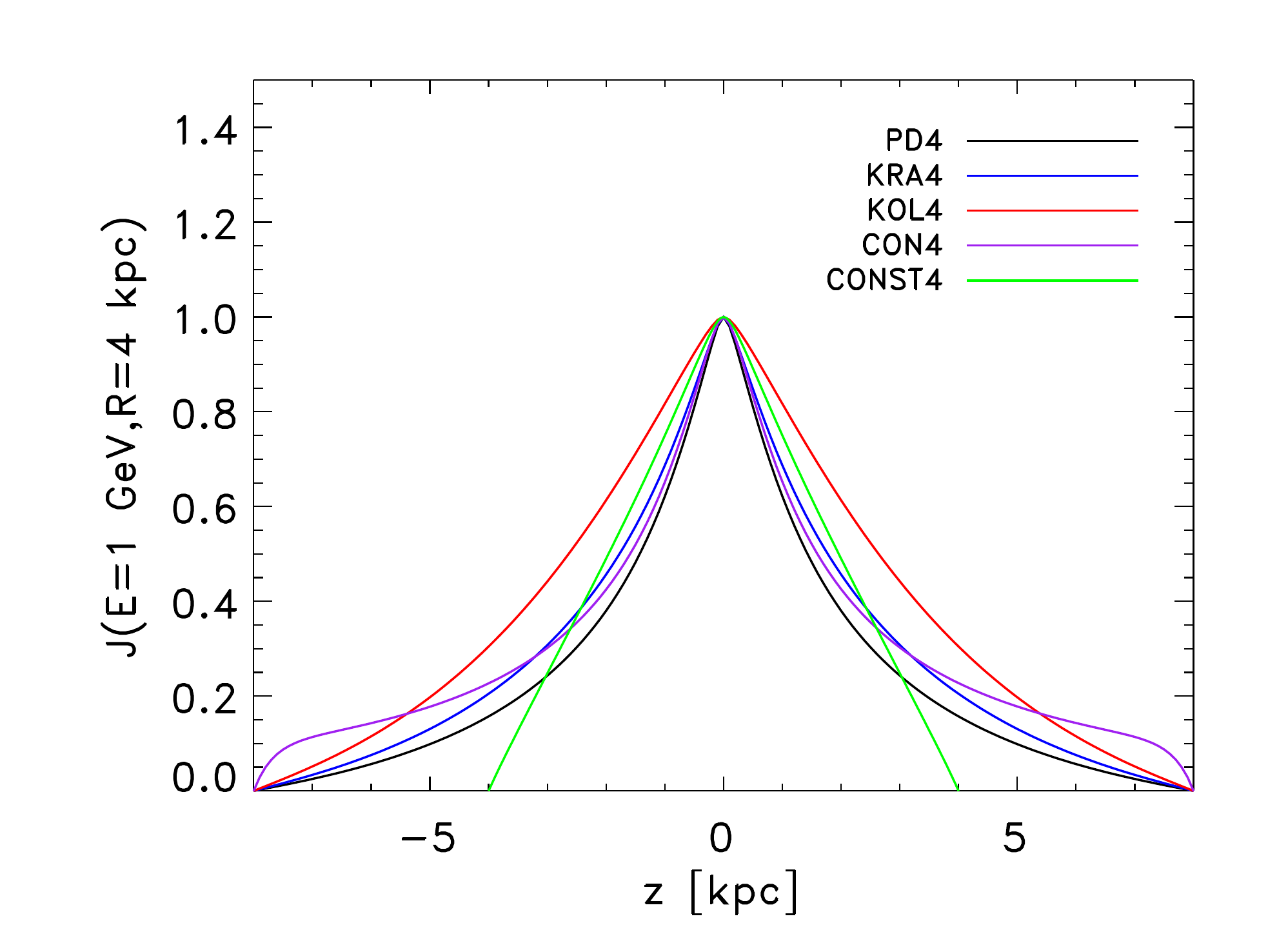}
\includegraphics[width=0.47\textwidth]{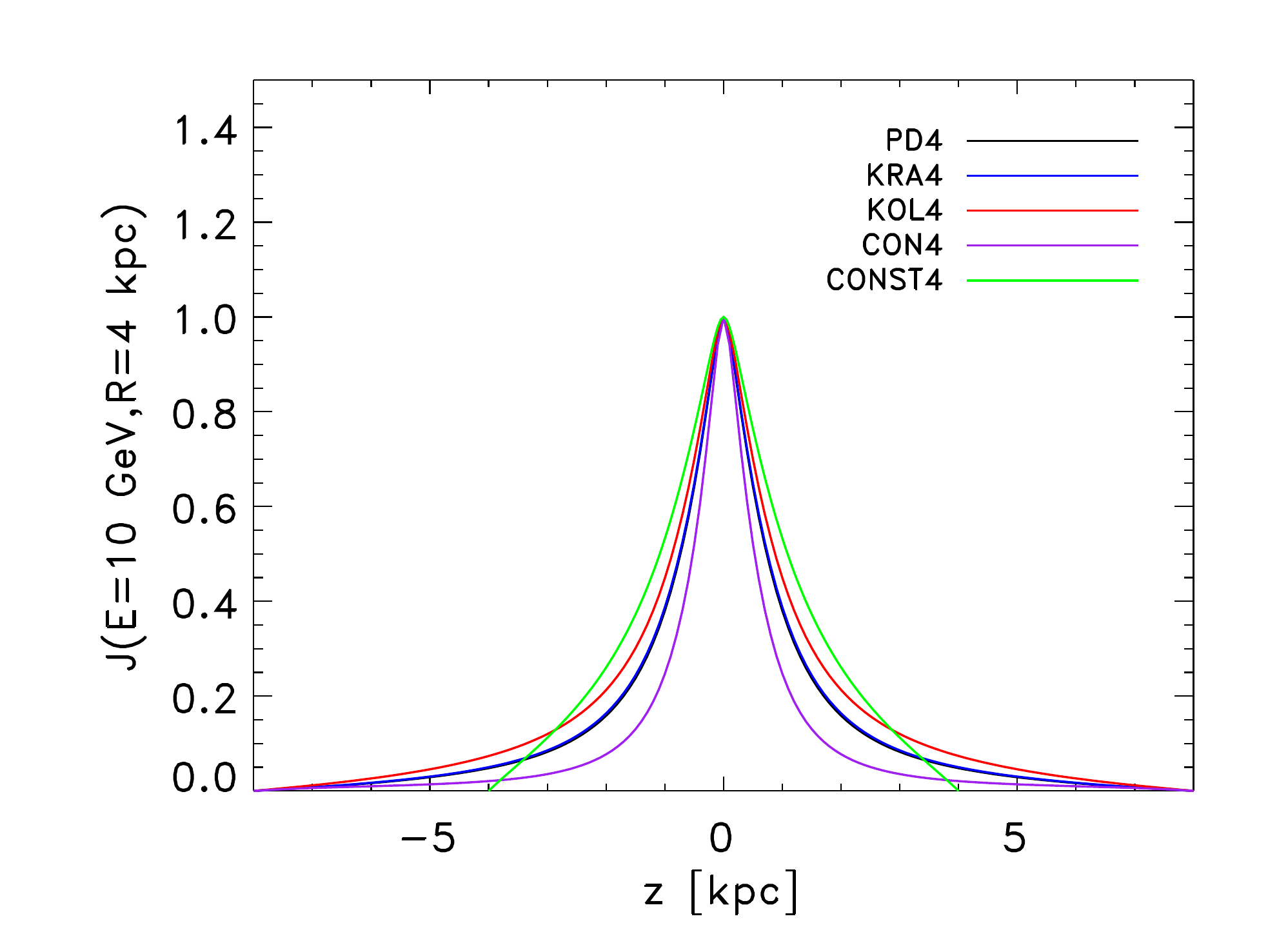}
\includegraphics[width=0.47\textwidth]{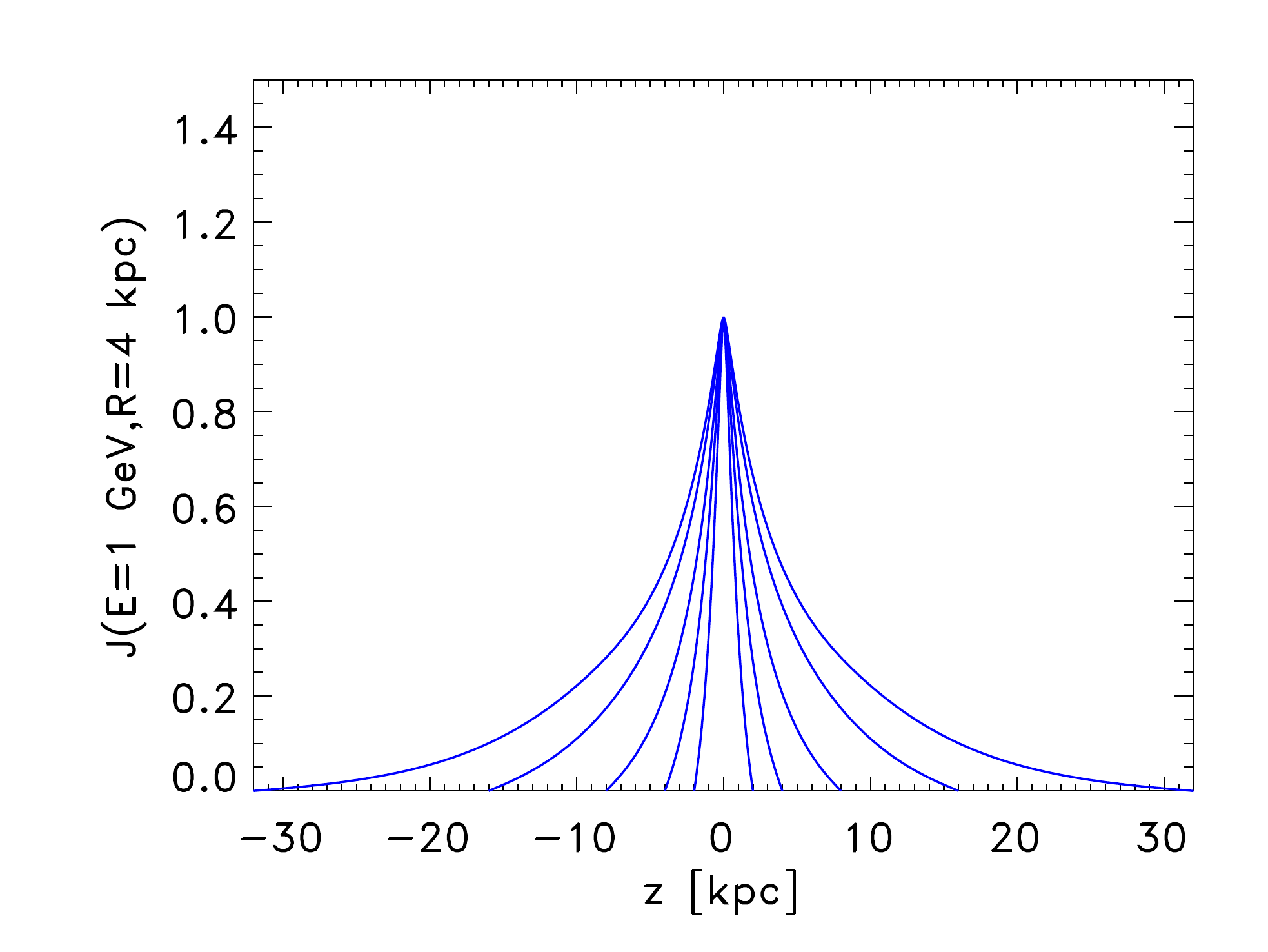}
\caption{{\it Top panels}: The vertical profile of the $e^-$ density at $R = 4~\kpc$ is plotted for $E = 1$ and $10~\GeV$ and for different propagation setups. 
The CONST4 (green lines) setup is similar to the KRA4  (and it also fit B/C and proton data), but the choice of a uniform diffusion coefficient and a random GMF up to $z_{\rm max} = z_t$. 
{\it Bottom panel}: the same quantity computed for the KRA setup with $z_t = 1,2,4,8,16~\kpc$ at $E = 1~\GeV$ . Fluxes are normalized to their values on the galactic plane at $z = 0$.}
\label{fig:CRE_profile_vs_zt} 
\end{figure}

Our first argument exploits the $z_t$-dependence of the synchrotron emission flux at medium and high Galactic latitudes. The flux observed in a given direction is proportional to the line of sight integral $ \int B^2(l) n_e(l) dl$, where $n_e$ is the $e^- + e^+$ CR density and the integral extends to the boundaries of the computational domain.  
In figure \ref{fig:CRE_profile_vs_zt} we show the $n_e(z)$ profile computed with \dragon\ for several propagation setups and different values of $z_t$. 
At  $\sim 1~\GeV$ energy losses have a minor effect on the CRE large scale distribution. In fact, at that energy they are dominated by IC scattering (with the exception of the GP region where bremsstrahlung dominates). The IC diffusive loss length is of the order of a few kpc on the GP and increases rapidly away from the disk.  We showed in the previous section, that $B_{\rm ran}$ is the dominant GMF component contributing to the synchrotron emissivity. Consequently, for a given propagation setup,  the synchrotron flux depends only on the random field normalization $B_{\rm ran}(0)$ and on $z_t$.  Independent measurements of the former quantity, such as the ones provided by Rotation Measures, offer, therefore, a valuable probe of the latter when combined with the synchrotron emission.    

\begin{figure}[tbp]
\centering
\includegraphics[width=0.6\textwidth]{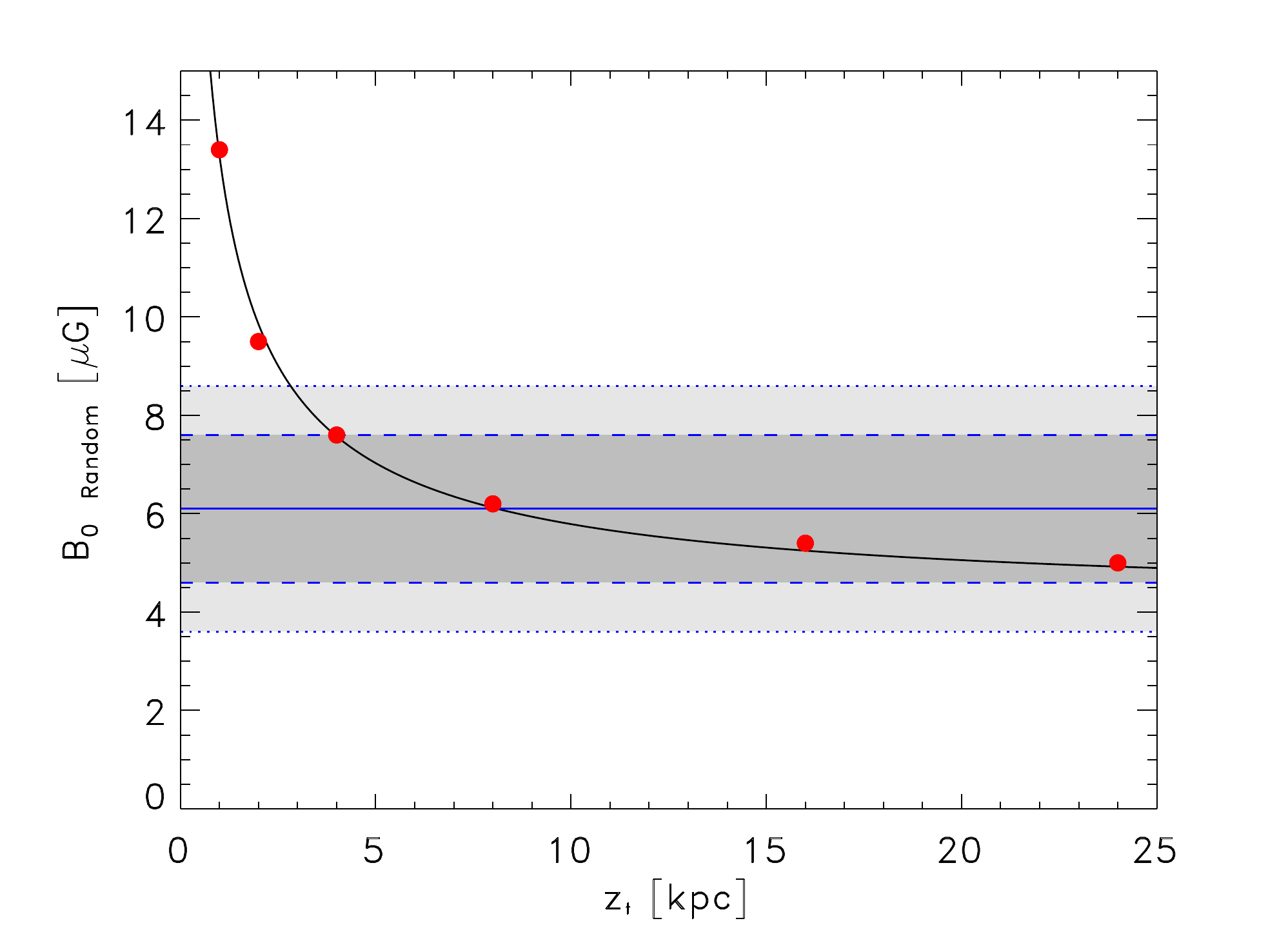}
\caption{The normalization of the random GMF  is plotted against its vertical scale height (which we assume to be the same as that of the diffusion coefficient).
The $3(5)~\sigma$ regions allowed by RM data are represented in gray (light-gray). Red dots are our results obtained under the condition that KRA models reproduce the observed synchrotron spectrum.  The black line is a $B_{\rm ran}^2 \propto 1/z_t$ fit of those points. The fitting function is $\left(B_{\rm ran}/1~\muG\right)^2 = 148.06\,(1~\kpc/z_{t})+19.12$. The fit computed for the other setups considered in this work would superimpose to that line.}
\label{fig:B_zt} 
\end{figure}

We consider then the synchrotron spectrum. For each value of $z_t$ we tune $B_{\rm ran}(0)$ so that the computed spectrum reproduces the data at 408 MHz.
Interestingly, we find a tight relation $B_{\rm ran}^2(0) \propto z_{t}^{-1}$ (see figure \ref{fig:B_zt}). 
This relation corresponds to the physical requirement that the total energy of the GMF be independent of the choice of $z_t$. 
In other words, the synchrotron spectrum considered here is sensitive to, and fixes, the total magnetic energy of the galaxy (see \cite{1996ARA&A..34..155B} for a similar discussion).

In figure \ref{fig:B_zt} we report the 3 and $5~\sigma$ regions allowed by RMs of Galactic pulsars \cite{Han:2004aa}  (see section \ref{sec:GMF}).  
Values of $z_t$ smaller than $4 (3)~ \kpc$ are excluded at 3 (5) $\sigma$. 
Thick CR halos, however, cannot be excluded on the basis of this argument. The plot in figure \ref{fig:B_zt} is computed for the KRA setup. We find no significant differences between this setup and the PD so that the same constraint applies to different propagation conditions. 

\begin{figure}[tbp]
\centering
\includegraphics[width=0.6\textwidth]{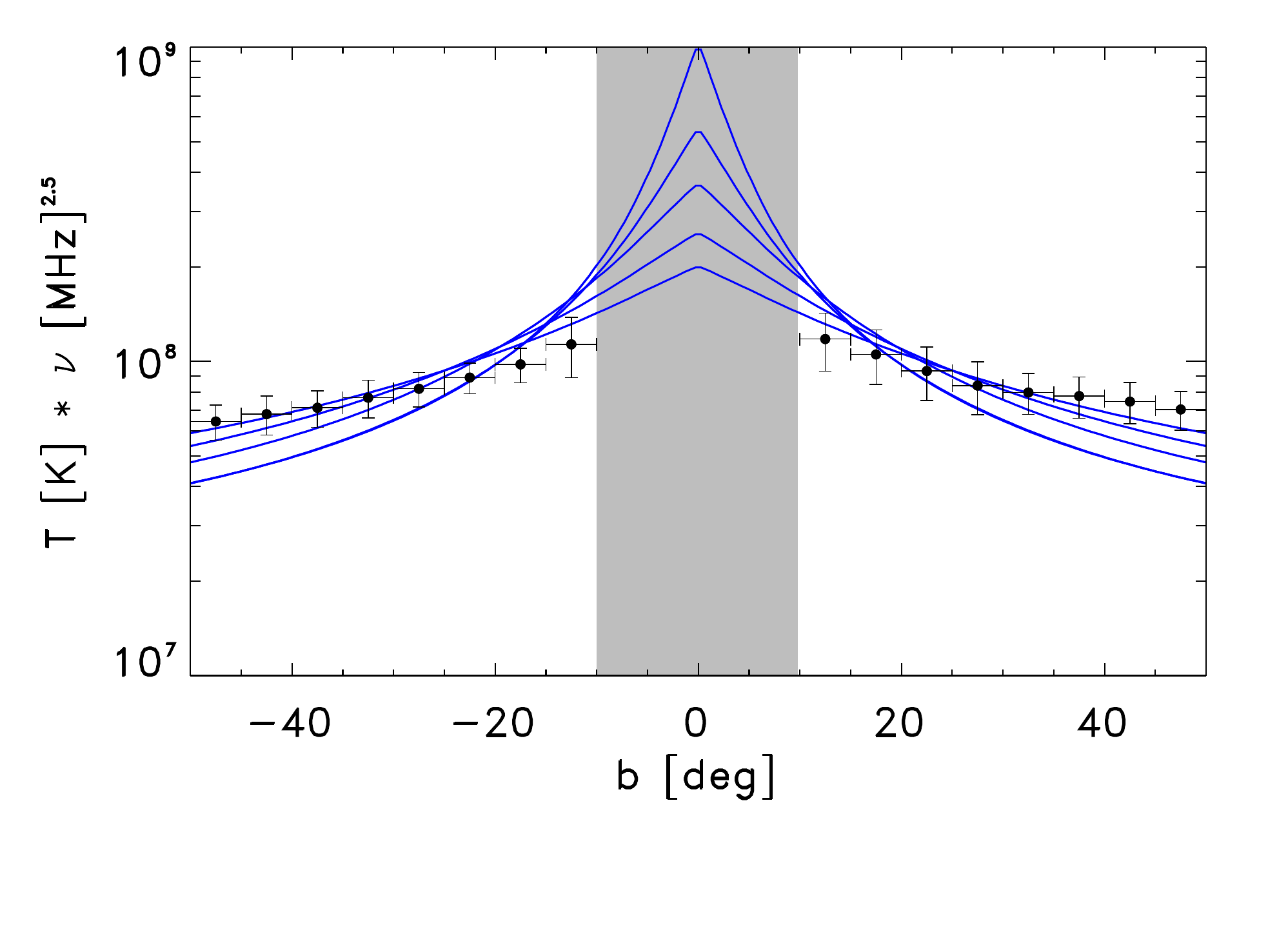}
\caption{The latitude profiles of the synchrotron emission at 408 MHz in the region 
$40^\circ < l < 100^\circ$ computed for the KRA propagation setup and $z_t = 1,2,4,8,16~\kpc$ are compared with radio data. The grey shadowed region is not considered when placing the constraint.}
\label{fig:lat_prof} 
\end{figure}

A complementary probe of the vertical distribution of CRE is offered by the latitude profile of the synchrotron emission. 
In figure \ref{fig:lat_prof} we compare the observed latitudinal profile of the synchrotron emission at 408 MHz to that calculated for the KRA setups for different values of $z_t$.        
As done before,  we look at the region $40^\circ < l < 340^\circ$, $10^\circ < b < 45^\circ$. 
Error bars correspond to the semi-dispersion of the distribution of the observed flux in those regions. We also consider the subregion $40^\circ < l < 100^\circ$, $10^\circ < b < 45^\circ$, where the flux variance (which is likely dominated by non-subtracted local structures) is smaller, which yields more stringent constraints.  
For each $z_{t}$ we tune the value of $B_{\rm ran}(0)$ so that the average spectrum in these regions is reproduced. We see that low values of $z_t$ are disfavored. 
This agrees with the results of the spectral analysis. A $\chi^{2}$ analysis shows that $z_{t}\leq 2~\kpc$ are excluded at the 3$\sigma$ level. The relevant values are reported in table \ref{tab:chi2}.
\begin{table}[tbp]
\begin{center}
\begin{tabular}{|c|c|}
\hline
$z_{t}~(\kpc)$ & $\chi^{2}$ \\
\hline
1 & 2.7 \\
2 & 2.5 \\
4 & 1.6 \\
8 & 0.9 \\
16 & 0.4 \\
\hline
\end{tabular}
\end{center}
\caption{Reduced $\chi^{2}$ values (13 degrees of freedom) of the comparison between synchrotron profiles and observational data in the region $40^\circ < l < 100^\circ$, $10^\circ < b < 45^\circ$.}
\label{tab:chi2}
\end{table}%

One of the most serious theoretical uncertainties which may affect the results of this subsection arises from the poorly known form of the vertical profile of the random GMF, hence of the diffusion coefficient. In order to quantify this uncertainty we repeated our analysis using a Gaussian profile (rather than the exponential used above) and, as an extreme possibility, 
a step function profile $f(z) = \theta(|z| - z_t)$  (which is the profile adopted in \cite{Strong:2011wd,Bringmann:2011py}) for both quantities. 
As a consequence, smaller field strengths are required on the GP to reproduce the synchrotron spectrum normalization and the curve in figure \ref{fig:B_zt} translates downwards. This would turn into a smaller lower limit on $z_t$, from $3$ to $2(1)~\kpc$ at $5~\sigma$  for the Gaussian (step-like) profile.  
We also find that $z_t > 9(10)~\kpc$ are excluded at $3 \sigma$ (but still allowed at $5 \sigma$).
We notice, however, that a step-like profile is physically hard to justify, as the gas density, which hosts the currents that should sustain the magnetic fields, decreases rapidly outside the galactic plane.  

\begin{figure}[tbp]
\centering
\includegraphics[width=0.6\textwidth]{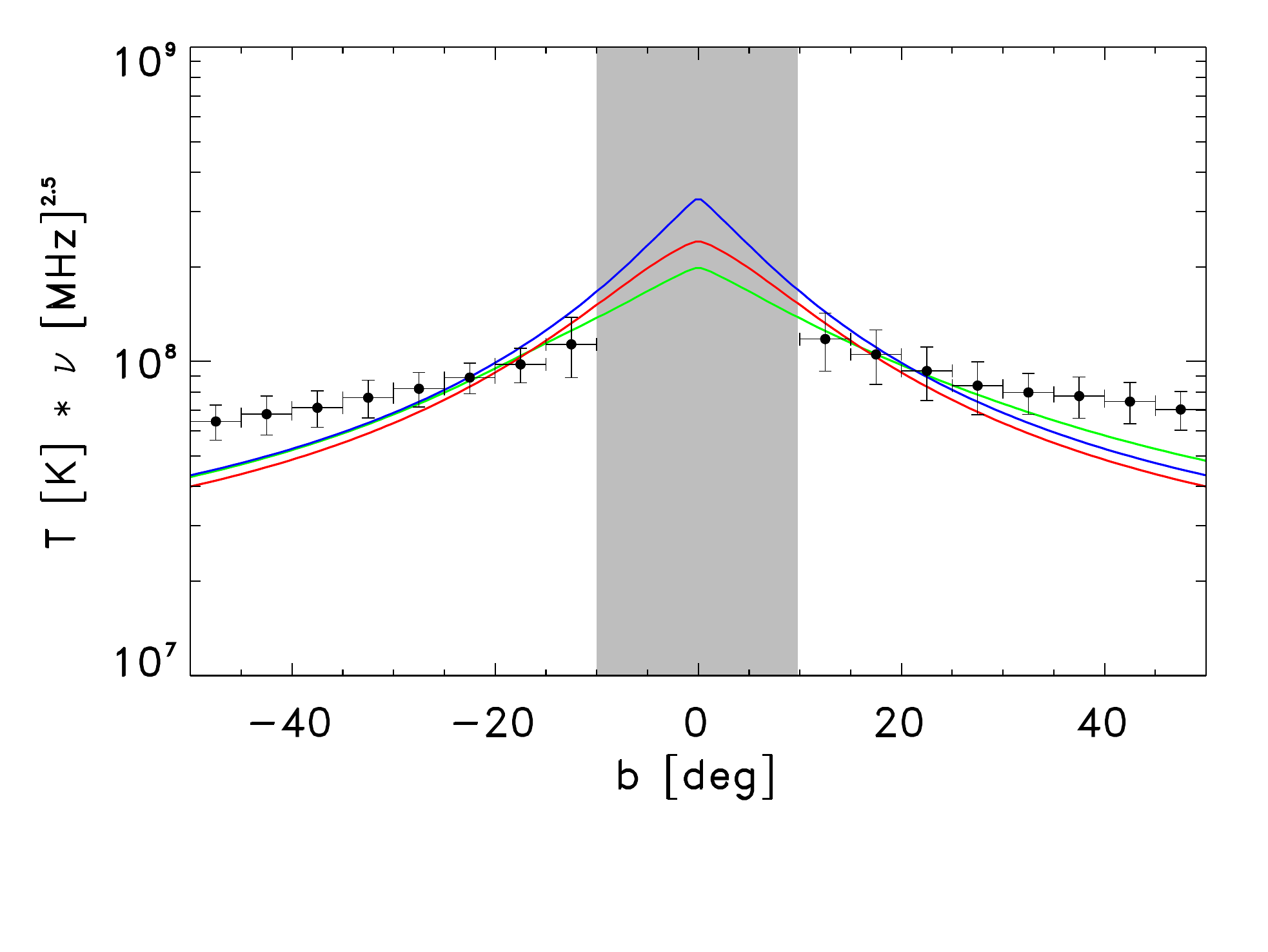}
\caption{The latitude profiles of the synchrotron emission at 408 MHz in the regions $40^\circ < l < 100^\circ$  computed for the KRA4 propagation setup with an exponential profile of the random component of the GMF (blue line) is compared with that obtained with similar models but a step-like profile (CONST4) (green line) and a Gaussian profile (red line). The grey shadowed region is the same as in figure \ref{fig:lat_prof}. }
\label{fig:kra_vs_const} 
\end{figure}

We also consider the effect of changing $f(z)$ on the latitude profile of the synchrotron emission.  
In Fig.\ref{fig:CRE_profile_vs_zt} we already compared the CRE profiles computed for the KRA4 with the CONST4, which adopts a step-like $f(z)$, finding small differences. In figure \ref{fig:kra_vs_const} we see as the latitude profile of the synchrotron emission obtained for those setups differ by $20\%$ at most
(the Gaussian profile which is also shown in that figure is even closer to the exponential one used for the KRA4 setups). 
Therefore, we conclude, that observable offers a reliable probe of the CR diffusive halo scale height.
 
\subsection{\boldmath The effects of varying $z_t$ on the electron and positron spectra}

In section \ref{subsec:cre_spectra} we showed that the low energy CR positron spectrum can be conveniently used to probe some of the CR propagation properties, in particular the effectiveness of reacceleration. Here we demonstrate that the same can be done also for $z_t$.

\begin{figure}[tbp] 
\centering
\subfigure[] {
\includegraphics[width=0.47\textwidth]{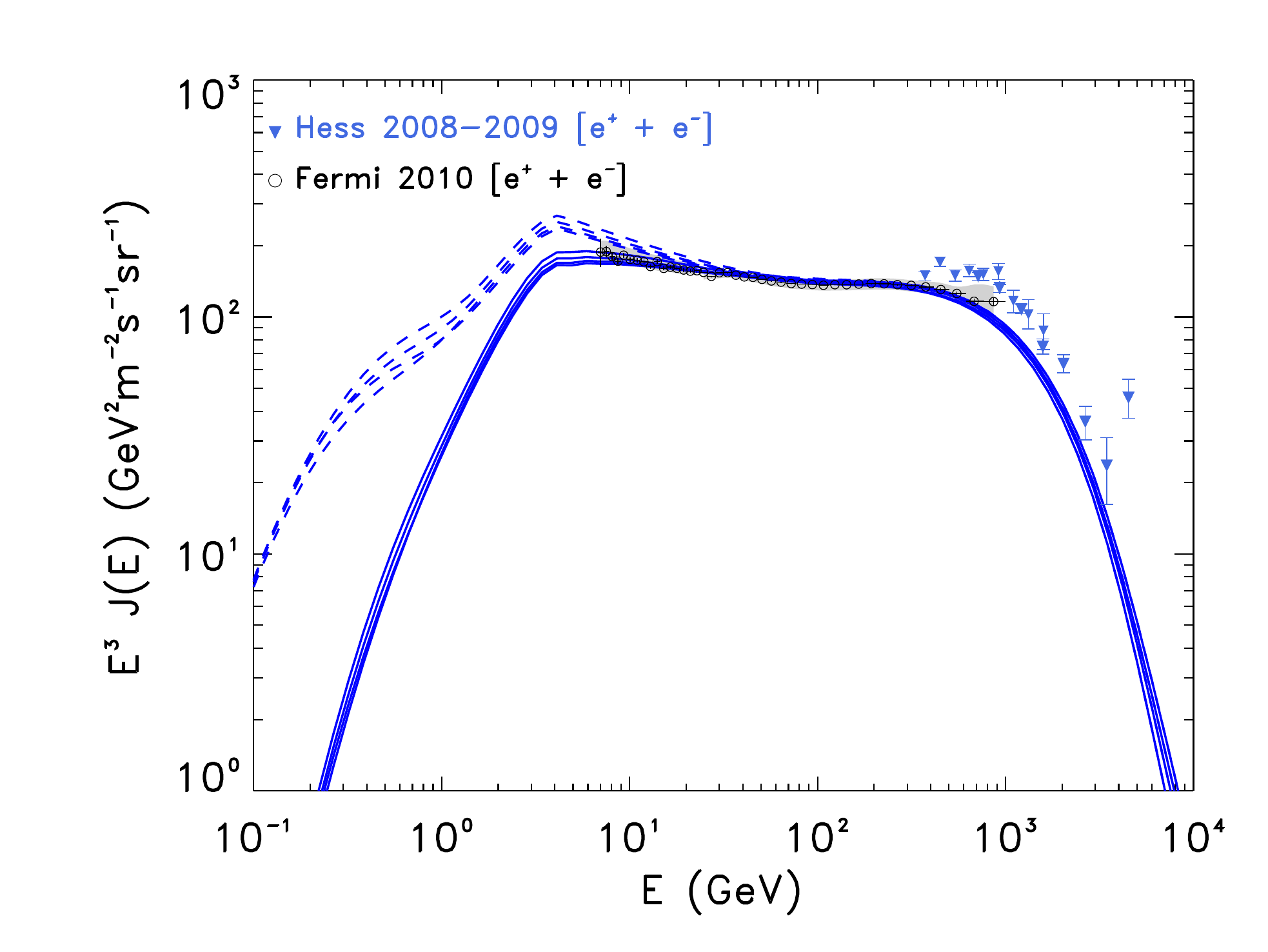}
}
\subfigure[] {
\includegraphics[width=0.47\textwidth]{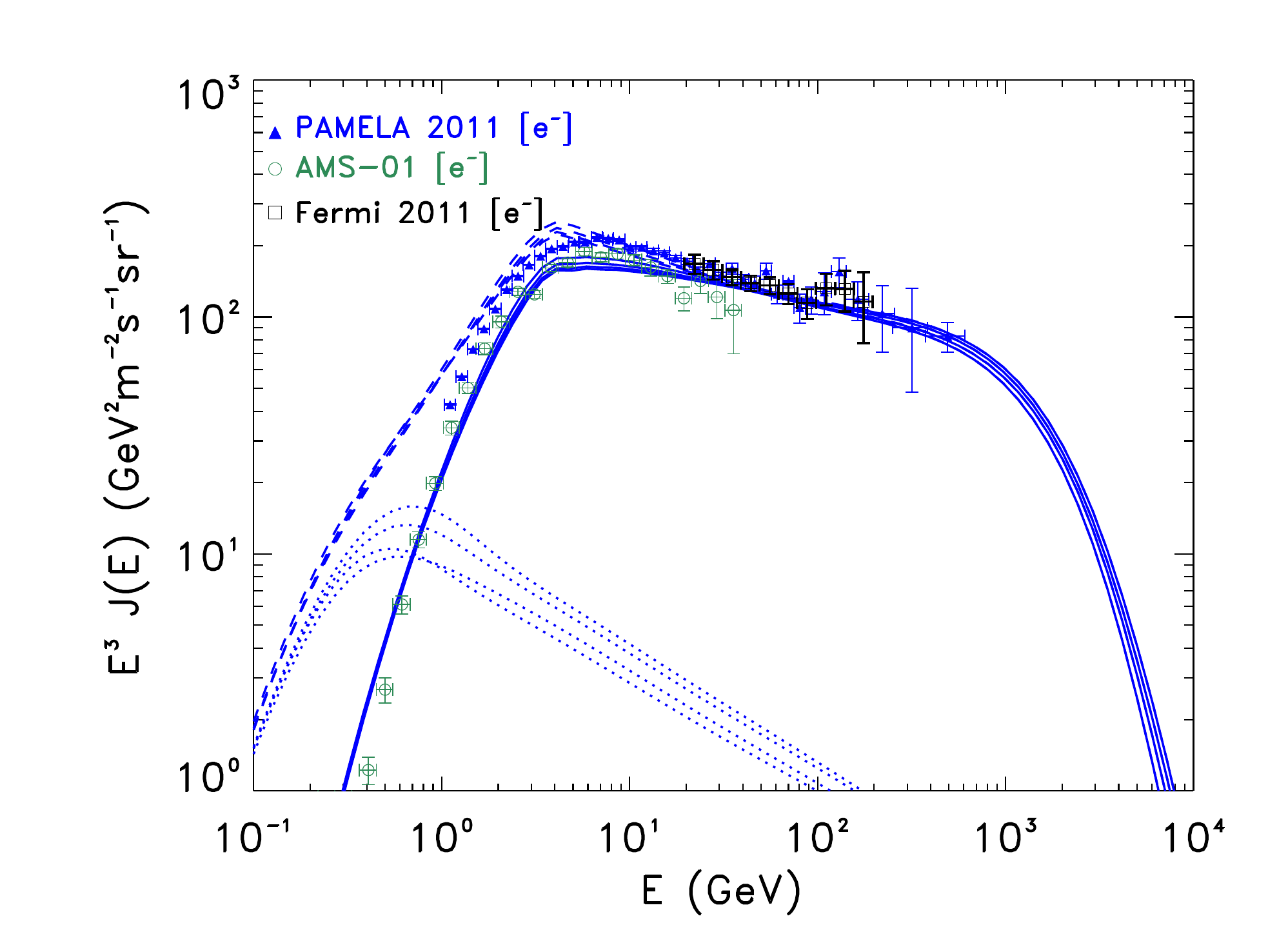}
}
\subfigure[] {
\includegraphics[width=0.47\textwidth]{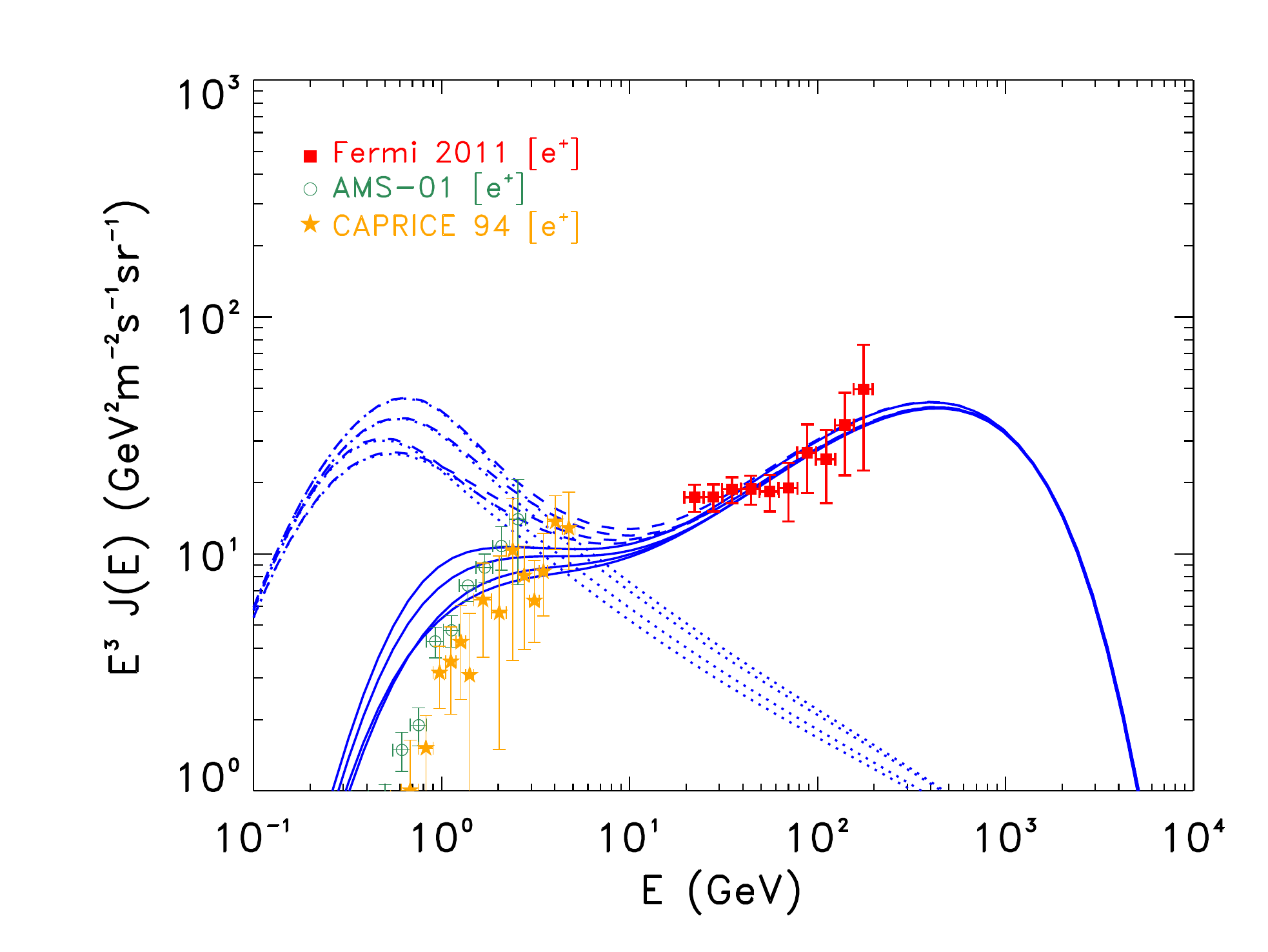}
}
\subfigure[] {
\includegraphics[width=0.47\textwidth]{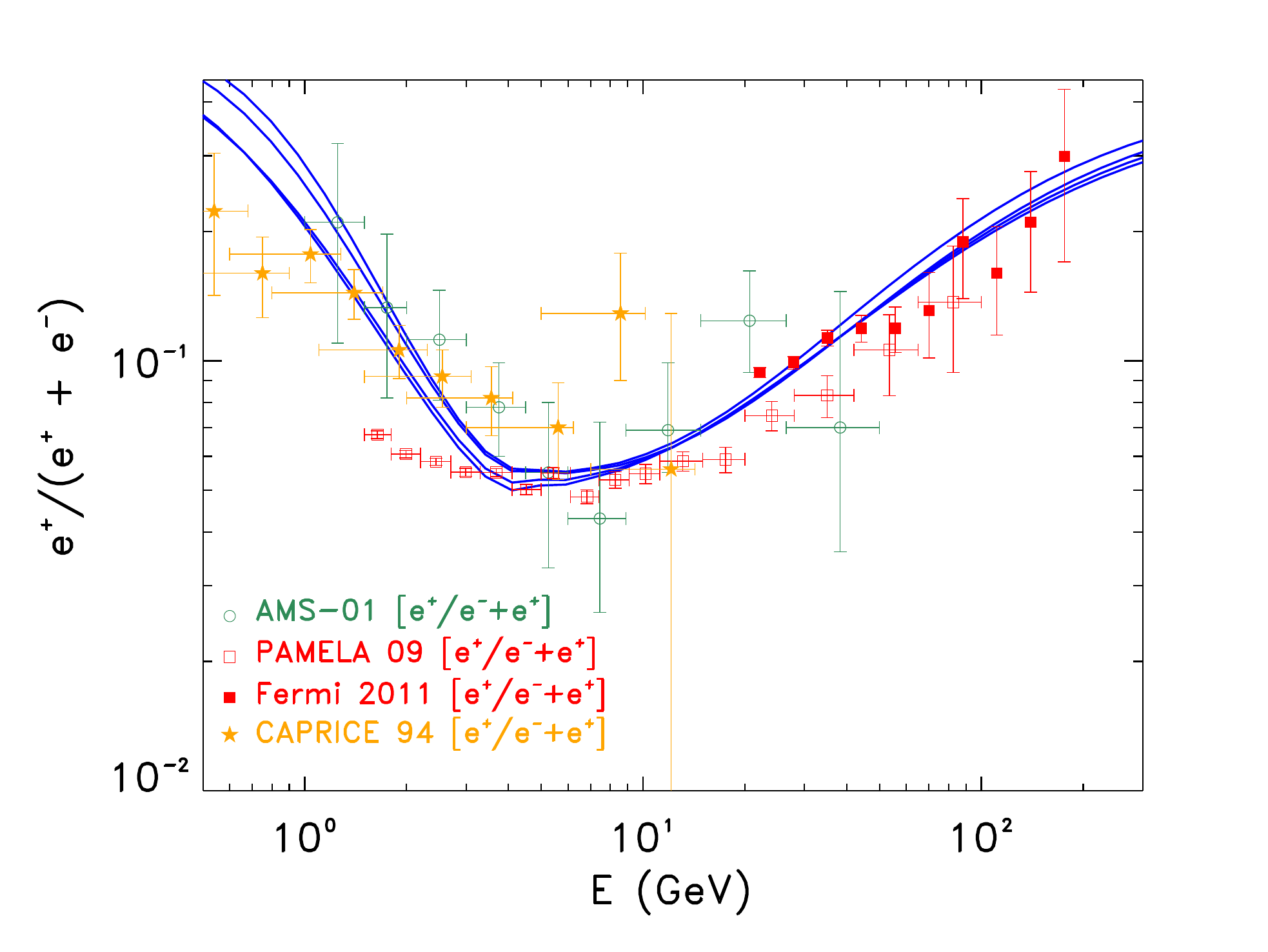}
}
\caption{The $e^+ + e^-$ (panel $a$), $e^-$ (panel $b$) and $e^+$ (panel $c$) spectra multiplied by $E^3$ as well as the positron fraction (panel d) 
spectra computed for the KRA setup and $z_t = 1,2,4,8~\kpc$ (from top to bottom respectively) are shown against a selection of experimental data. The line style notation is the same as in figure \ref{fig:models_FERMI}.  All models are modulated with $\Phi = 0.3~{\rm GV}$. 
}
\label{fig:crspectra_vs_zt}
\end{figure}

We remind the reader that the B/C, or other secondary-to-primary ratios, can constrain only the ratio $D_0/z_t$, where $D_0$ is the normalization of the diffusion coefficient (see eq.~\ref {eq:diff_coeff}). The secondary $e^+$  flux however depends significantly on the scale height $z_t$, thus breaking the degeneracy of this parameter with $D_0$. This can be promptly understood from the simple general equation (see e.g.~\cite{Bulanov_1974Ap&SS})
\begin{equation}
N(e) = \frac{ \tau~ Q~ V_S}{V}     
\end{equation}
where $\tau$ is the minimum between the escape and loss times, $Q$ the production rate per unit volume, $V_S$ and $V$ the source and CR filling volumes respectively.
For secondary electrons and positrons $Q \propto \sigma_{pp} n_{\rm gas} N_p$ where $N_p$ is the CR proton density in the GP.  
We showed that at the most relevant energy $E \simeq 1~\GeV$ the CRE fill the entire volume occupied by the random GMF, so that $V \propto z_t$. 
This also implies that energy losses are subdominat and that $\tau $ is determined by the propagation setup through the $D_0/z_t$ ratio (thus $\tau_{\rm esc}$). 
As a consequence, for a given setup the secondary electron and positron densities decrease with increasing $z_t$. 
Since this is a dilution effect, it is expected to be energy independent which is what we find from our numerical computations (see figure \ref{fig:crspectra_vs_zt}).
This effect is most clearly detectable in the $e^+$  channel since below 10 GeV  the contribution of primary positrons from the extra-component is negligible.
Figure \ref{fig:crspectra_vs_zt} shows that the $e^+$ spectrum measured by AMS-01 disfavors values of $z_t$ smaller than $2~ \kpc$. 
This strengthens the conclusions reached in the first part of this section on the basis of radio observations. 
In figure \ref{fig:crspectra_vs_zt} we only show spectra computed with the KRA models since the other setups yield similar results. Remarkably, even large values of $z_t$ cannot reconcile strong reacceleration models with the positron data, because the dilution effect cannot compensate for their typical bump at $\sim1~\GeV$. 

\section{Discussion and conclusions}\label{sec:discussion}

One of the main implications of this work is that below a few GeV the injected electron spectrum must be drastically suppressed with respect to a simple power-law extrapolation of the one which fits CRE Fermi-LAT data.  This suppression can be obtained by introducing either a break or an exponential IR cutoff in the $e^-$ source spectrum below a few GeV. This last possibility is particularly intriguing, as it might hint at a suppression of the leakage of accelerated electrons from the sources at GeV and sub-GeV energies.  Furthermore, similarly to \cite{Strong:2011wd}, we find a rather high source spectral index $\gamma_0(e^-) \simeq 2.5$ above a few GeV.  This is larger than the value inferred from SNR radio catalogues derived in \cite{Delahaye:2010ji} which is $\gamma_{\rm SNR} = 2.0 \pm 0.3$, albeit the two are compatible within $2\sigma$. As discussed in \cite{Strong:2011wd}, this may imply that the electron population probed by the SNR radio observations is not representative of that injected in the ISM. However, we point out that this has to be considered as an {\it effective} index, since the code does not take into account the presence of structures in the source distribution. Accounting for the spiral arm distribution of SNRs may result in a different requirement for the injection index: indeed, being the Sun in the so-called {\it local spur} situated in a interarm region, the average distance from SNRs is larger than in the smooth case: as a consequence,
a harder injection spectrum may be required to compensate for the larger energy losses and reproduce the observed spectrum. 

We find a break amplitude similar to that in \cite{Bringmann:2011py}.  In that paper, however, the break was imposed on the total (primary + secondary) propagated spectrum.  It was claimed that leaving the secondary spectra unaffected would lead to overproduction of synchrotron radiation. Our figure \ref{fig:synchro_vs_setups}, however, shows that this is not the case.  Below $\sim 100$ MHz the contribution of secondary electrons and positrons becomes gradually sufficient to account for the observed radio spectrum and no break is required in the spectra of these particles.   For opposite reasons, we disagree with the results of \cite{Strong:2011wd}, as in that paper it was found that secondary leptons can contribute at most one-third of the low energy synchrotron spectrum.  This difference can be explained by the break in the proton and Helium injection spectra in their models \cite{Strong:2010pr}, which suppresses secondary production at low energy. We remark that we adopt a break only in KOL and CON models, while in the other, experimentally favorite models the primary nuclei spectra are a single power-law in momentum.

A relevant consequence of our result is that an interpretation of the low energy (low frequency) behavior of the CRE (synchrotron) spectrum in terms of a break in the rigidity dependence of the diffusion coefficient (which should affect all components) is disfavored. 
A strong break, as proposed in \cite{Webber_2008JGRA}, is hardly compatible with both CR nuclei and lepton spectra (a multichannel analysis is not provided in that paper). 
   
One relevant aspect of our approach is that it is based on a multichannel analysis of nuclear, electron and positron spectra. Positrons are especially important because their low energy spectrum is dominated by secondary particles. We find that once the low energy  $e^-$ source spectrum is tuned to reproduce the observed synchrotron spectrum, only models featuring low reacceleration can reproduce the observed $e^+$ spectrum and fraction. This conclusion is in agreement with \cite{Strong:2011wd} and it strengthens the results of our previous analyses based on antiprotons \cite{DiBernardo:2009ku} and $e^+ + e^-$ spectra \cite{DiBernardo:2010is}.  

Our modeling of the synchrotron emission of the Galaxy accounts, for the first time in this framework, for the presence of the $e^\pm$ extra-component. This is required not only to consistently model PAMELA and Fermi-LAT high energy data ($E \gg 10~\GeV$) but also to correctly estimate the $e^-$ source spectrum from CR and radio data. In this work we treated the extra-component source term as a continuous distribution and assumed it to spatially coincide with that of the standard $e^-$ component.  This may be not realistic, because the extra-component could well be powered by one or a few nearby sources \cite{Grasso:2009ma}. We find however that adopting the latter hypothesis will not affect significantly any of our results.  Indeed, we have shown in Fig.~\ref{fig:Ecut_spectrum} that electrons and positrons with $E > 20\GeV$ (where the extra-component starts playing a role) 
give a negligible contribution to the synchrotron spectrum in the explored frequency range.
As a consequence, subtracting the contribution of the extra-component from the  $e^+ + e^-$ Galactic spectrum does not change the synchrotron spectrum at a level detectable with the presently available surveys.  
However, by combining high precision CRE data from AMS-02 \citep{Incagli:2010} and CALET \citep{Torii:2008} as well as forthcoming radio observations by PLANCK\footnote{\url{http://planck.esa.it}}, LOFAR\footnote{\url{http://www.lofar.org}} and SKA\footnote{\url{http://www.skatelescope.org}} can become possible to detect the contribution of the extra-component to the synchrotron spectrum. 

With our method we exploit the galactic diffuse synchrotron emission as a way to measure the low energy LIS spectrum of CR electrons and positrons, in a similar way as the diffuse $\gamma$-ray emission is being exploited to obtain the LIS proton spectrum \cite{Neronov:2011wi,Dermer:2012bz,Kachelriess:2012fz}. This is a valuable information for studies of solar modulation. While the simple force-field approach works quite well in reproducing AMS-01 data, PAMELA data require a more detailed investigation of charge-sign dependent effects in the solar system.

\begin{figure}[tbp]
\begin{center}
\includegraphics[width=0.45\textwidth]{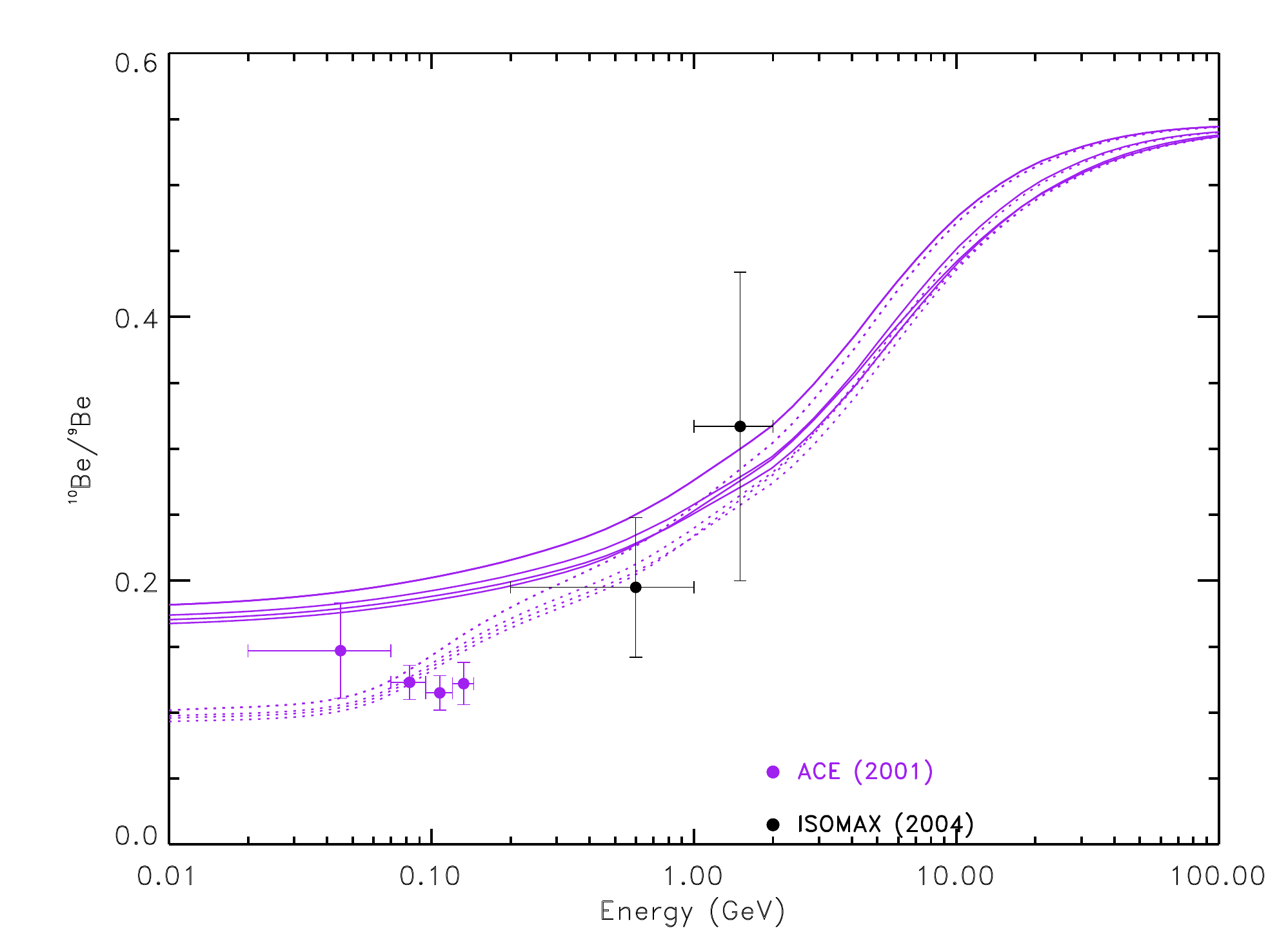}
\includegraphics[width=0.45\textwidth]{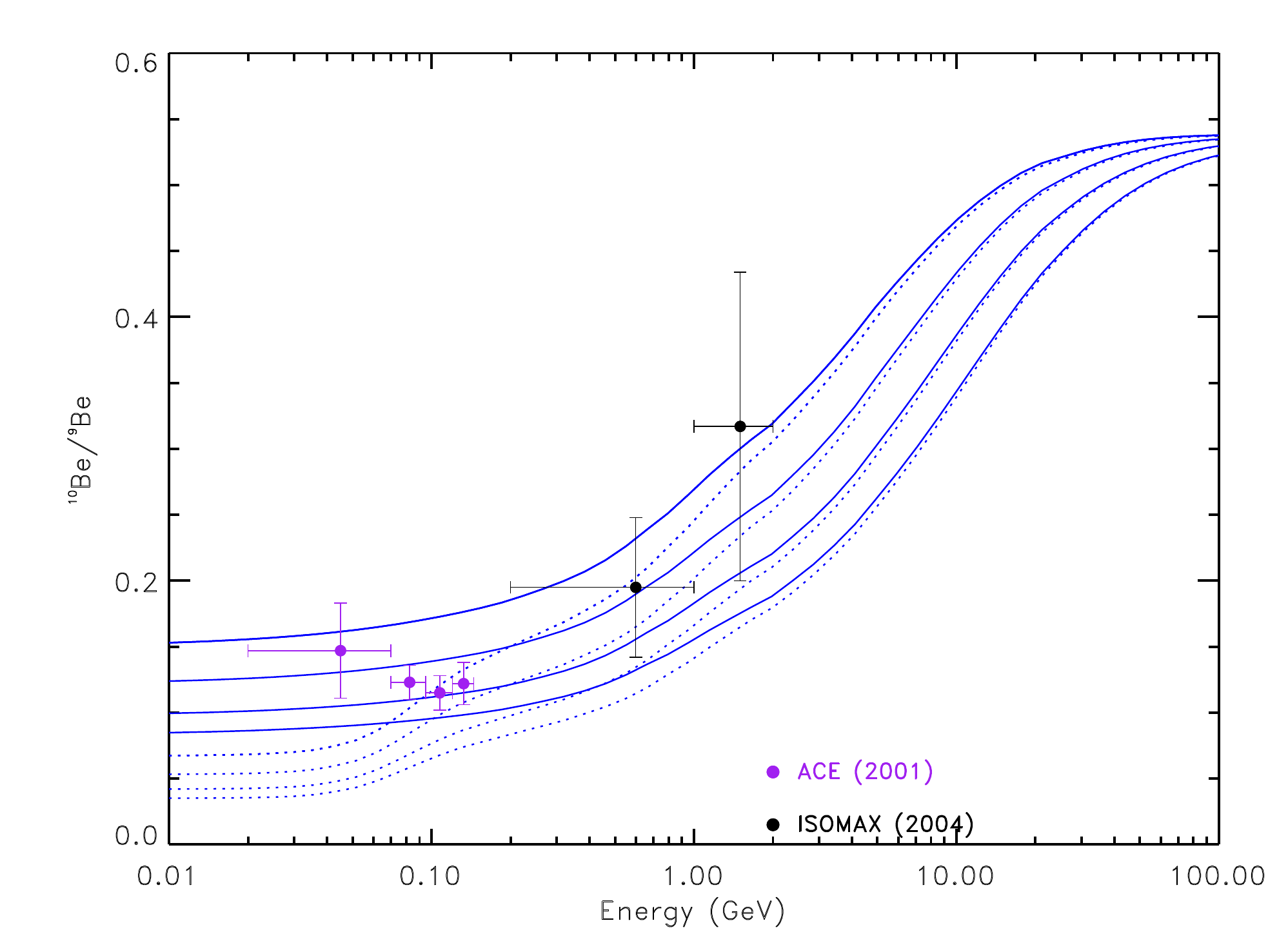}
\caption{$^{10}{\rm Be}/^{9}$Be modulated (dotted) and non-modulated (solid) spectra for the KRA (left panel) and CON (right panel) configurations. From bottom to top curves correspond to $z_{t} = 1, 2, 4, 8~\kpc$.}
\label{fig:Be}
\end{center}
\end{figure}

We investigate the vertical extension of the CRE distribution by means of several complementary methods.
Remarkably, these methods are more powerful than the $^{10}{\rm Be}/^{9}$Be in placing lower limits on $z_{t}$ and also more robust because:
\begin{itemize}
\item They come from non-local observables and they are not sensitive to the effects of solar modulation or local physics (e.g.~the possible presence of a local bubble \cite{Donato:2001eq}); 
\item the constraints based on $^{10}{\rm Be}/^{9}$Be strongly depend on the propagation model.
\end{itemize}
Given the importance of these points, we show for reference in figure \ref{fig:Be} the vertical scale dependence of the $^{10}{\rm Be}/^{9}$Be in our KRA and CON models, together with available experimental data from ACE and ISOMAX. The most discriminatory points are from ACE where they are susceptible to solar modulation.
The dependence on the propagation setup is clearly seen by comparing the KRA and the CON model. Qualitatively, the difference between the two cases is that in convective models the convective outflows along the normal to the disc plane
introduce an effective vertical scale height \cite{2001ApJ...555..585M,Taillet:2002ub}. This explains why in the CON case the $^{10}{\rm Be}/^{9}$Be is less sensitive to $z_t$  than in the KRA case. As a consequence, the constraining power of $^{10}{\rm Be}/^{9}$Be is reduced for convective models. The results we obtained from our new analysis are not plagued by such uncertainties.

Having lower limits on the halo scale height is of great importance for indirect DM searches. Because the sources of DM originated particles are distributed also in the halo (in contrast with astrophysical sources that are more localized in the disk), the predicted fluxes depend strongly on the extension of the halo. In particular, constraints on the annihilation cross section are considerably weakened if the halo size is small. According to \cite{Evoli:2011id}, imposing the limit $z_{t}>0.5~\kpc$ would improve the constraints  on the annihilation cross section by a factor of two to ten.
A full account of the consequences of our bound $z_{t}>2~\kpc$ is beyond the scope of this paper and will be left for a forthcoming work.

Our lower limit on $z_t$ may have also important implications for the interpretation of the CR anisotropy. In fact, it was shown that increasing the diffusive halos scale height reduces the possible role of source stochasticity to explain the large scale anisotropy fluctuations observed above 10 TeV~\cite{Blasi:2012}.    

For the first time, we have placed a constraint on the CR diffusive halo scale height based on the comparison of the computed synchrotron emission intensity with observations. It is based on the CRE spectra measured by Fermi-LAT and GMF constraints determined with Faraday RMs.  Our constraint depends weakly on the poorly known vertical profile of the random component of the GMF being $z_t \le 3(2)~\kpc$ for a exponential (Gaussian) profile. 
Statistical and systematic errors on $B_{\rm ran}$ are expected to be significantly reduced in the near future by LOFAR and SKA. 
Similarly to what was done in \cite{Bringmann:2011py} we also compare the computed latitude profile with radio maps of selected regions, obtaining consistent results and excluding $z_{t}\leq2~\kpc$ at 3$\sigma$ confidence level. We remark that constraints derived in this way are stronger than the ones obtained from $\gamma$-ray data \cite{Cholis:2011un}.

Differently from previous works which assumed a uniform diffusion coefficient, we adopted here a vertical dependence $D(z)$ correlated with that of the turbulent magnetic field. We found that this turns into a different distribution of the electron density which, however, does not affect the latitude profile of the synchrotron emission at a level detectable with the available surveys (see Fig.~\ref{fig:kra_vs_const}).  
We also considered the possibility that the diffusion coefficient has a radial dependence $D(r)$ correlated with magnetic field radial profile or the CR source distribution as proposed to provide a combined solution of the $\gamma$-ray gradient and cosmic ray anisotropy problems \cite{Evoli:2012ha}. 
We found no significant effect on the synchrotron spectrum. The effect on the emission morphology, which may be sizable, will be studied in a forthcoming paper.

Finally, we used the low energy positron spectrum and positron fraction dependence on $z_t$ to constrain this quantity. Also this method seriously disfavors low values of $z_t$. AMS-02 data and recent progresses in solar modulation modeling should soon considerably improve the usefulness of this approach.  
  
\acknowledgments
We are grateful to Iris Gebauer, Wim de Boer and the IEKP for providing powerful computing resources. 
We also thank F.~Donato, J.~Han, P.~D.~Serpico, A.~Strong and U.~Torkelsson for useful comments and suggestions.
CE~acknowledges support from the Helmholtz Alliance for Astro-particle Physics funded by the Initiative and Networking Fund of the Helmholtz Association.  D.~Gaggero warmly thanks the Institute of Experimental nuclear Physics (IEKP) of the Karlsruhe Institut f\"ur Technologie (KIT) and the Max-Planck-Institut f\"ur Physik in M\"unchen for hosting and supporting him during the realization of this project. 
D. Grasso thanks the Institut f\"ur Theoretische Physik at Universit\"at Hamburg for warm hospitality and the Deutsche Forschungsgemeinschaft through the collaborative research centre SFB 676 for financial support.  LM~acknowledges support from the Alexander von Humboldt Foundation and partial support from the  European Union FP7 ITN INVISIBLES (Marie Curie Actions, PITN- GA-2011- 289442).
Some of the results in this paper have been derived using the {\tt HEALPix} \cite{2005ApJ...622..759G} package. 

\bibliography{synchrotron_paper}

\providecommand{\href}[2]{#2}\begingroup\raggedright\begin{thebibliography}{10}

\bibitem{Daniele}
D.~Gaggero, {\em Cosmic Ray Diffusion in the Galaxy and Diffuse Gamma
  Emission}.
\newblock Springer Theses, 2012.

\bibitem{Adriani:2008zr}
{\bf PAMELA Collaboration} Collaboration, O.~Adriani et~al., {\it {An anomalous
  positron abundance in cosmic rays with energies 1.5-100 GeV}},  {\em Nature}
  {\bf 458} (2009) 607--609, [\href{http://xxx.lanl.gov/abs/0810.4995}{{\tt
  arXiv:0810.4995}}].

\bibitem{Abdo:2009zk}
{\bf Fermi LAT Collaboration} Collaboration, A.~A. Abdo et~al., {\it
  {Measurement of the Cosmic Ray e+ plus e- spectrum from 20 GeV to 1 TeV with
  the Fermi Large Area Telescope}},  {\em Phys.Rev.Lett.} {\bf 102} (2009)
  181101, [\href{http://xxx.lanl.gov/abs/0905.0025}{{\tt arXiv:0905.0025}}].

\bibitem{Ackermann:2010ij}
{\bf Fermi LAT Collaboration} Collaboration, M.~Ackermann et~al., {\it {Fermi
  LAT observations of cosmic-ray electrons from 7 GeV to 1 TeV}},  {\em
  Phys.Rev.} {\bf D82} (2010) 092004,
  [\href{http://xxx.lanl.gov/abs/1008.3999}{{\tt arXiv:1008.3999}}].

\bibitem{Adriani:2011xv}
{\bf PAMELA Collaboration} Collaboration, O.~Adriani et~al., {\it {The
  cosmic-ray electron flux measured by the PAMELA experiment between 1 and 625
  GeV}},  {\em Phys.Rev.Lett.} {\bf 106} (2011) 201101,
  [\href{http://xxx.lanl.gov/abs/1103.2880}{{\tt arXiv:1103.2880}}].

\bibitem{Aguilar:2007yf}
{\bf AMS-01 Collaboration} Collaboration, M.~Aguilar et~al., {\it {Cosmic-ray
  positron fraction measurement from 1 to 30-GeV with AMS-01}},  {\em
  Phys.Lett.} {\bf B646} (2007) 145--154,
  [\href{http://xxx.lanl.gov/abs/astro-ph/0703154}{{\tt astro-ph/0703154}}].

\bibitem{Aharonian:2009ah}
{\bf H.E.S.S. Collaboration} Collaboration, F.~Aharonian et~al., {\it {Probing
  the ATIC peak in the cosmic-ray electron spectrum with H.E.S.S}},  {\em
  Astron.Astrophys.} {\bf 508} (2009) 561,
  [\href{http://xxx.lanl.gov/abs/0905.0105}{{\tt arXiv:0905.0105}}].

\bibitem{Grasso:2009ma}
{\bf FERMI-LAT Collaboration} Collaboration, D.~Grasso et~al., {\it {On
  possible interpretations of the high energy electron-positron spectrum
  measured by the Fermi Large Area Telescope}},  {\em Astropart.Phys.} {\bf 32}
  (2009) 140--151, [\href{http://xxx.lanl.gov/abs/0905.0636}{{\tt
  arXiv:0905.0636}}].

\bibitem{FermiLAT:2011ab}
{\bf Fermi LAT Collaboration} Collaboration, M.~Ackermann et~al., {\it
  {Measurement of separate cosmic-ray electron and positron spectra with the
  Fermi Large Area Telescope}},  {\em Phys.Rev.Lett.} {\bf 108} (2012) 011103,
  [\href{http://xxx.lanl.gov/abs/1109.0521}{{\tt arXiv:1109.0521}}].

\bibitem{Grasso:2011wt}
{\bf Fermi-LAT Collaboration} Collaboration, D.~Grasso and D.~Gaggero, {\it
  {Cosmic ray models compared to Fermi-LAT positron and electron separate
  spectra}},  \href{http://xxx.lanl.gov/abs/1110.2591}{{\tt arXiv:1110.2591}}.

\bibitem{Strong:2007nh}
A.~W. Strong, I.~V. Moskalenko, and V.~S. Ptuskin, {\it {Cosmic-ray propagation
  and interactions in the Galaxy}},  {\em Ann.Rev.Nucl.Part.Sci.} {\bf 57}
  (2007) 285--327, [\href{http://xxx.lanl.gov/abs/astro-ph/0701517}{{\tt
  astro-ph/0701517}}].

\bibitem{Trotta:2010mx}
R.~Trotta, G.~Johannesson, I.~Moskalenko, T.~Porter, R.~R. de~Austri, et~al.,
  {\it {Constraints on cosmic-ray propagation models from a global Bayesian
  analysis}},  {\em Astrophys.J.} {\bf 729} (2011) 106,
  [\href{http://xxx.lanl.gov/abs/1011.0037}{{\tt arXiv:1011.0037}}].

\bibitem{Putze:2010zn}
A.~Putze, L.~Derome, and D.~Maurin, {\it {A Markov Chain Monte Carlo technique
  to sample transport and source parameters of Galactic cosmic rays: II.
  Results for the diffusion model combining B/C and radioactive nuclei}},  {\em
  Astron.Astrophys.} {\bf 516} (2010) A66,
  [\href{http://xxx.lanl.gov/abs/1001.0551}{{\tt arXiv:1001.0551}}].

\bibitem{Evoli:2011id}
C.~Evoli, I.~Cholis, D.~Grasso, L.~Maccione, and P.~Ullio, {\it {Antiprotons
  from dark matter annihilation in the Galaxy: astrophysical uncertainties}},
  {\em Phys.Rev.} {\bf D85} (2012) 123511,
  [\href{http://xxx.lanl.gov/abs/1108.0664}{{\tt arXiv:1108.0664}}].

\bibitem{Evoli:2008dv}
C.~Evoli, D.~Gaggero, D.~Grasso, and L.~Maccione, {\it {Cosmic-Ray Nuclei,
  Antiprotons and Gamma-rays in the Galaxy: a New Diffusion Model}},  {\em
  JCAP} {\bf 0810} (2008) 018, [\href{http://xxx.lanl.gov/abs/0807.4730}{{\tt
  arXiv:0807.4730}}].

\bibitem{Han:2009ts}
J.~Han, {\it {Magnetic structure of our Galaxy: A review of observations}},
  \href{http://xxx.lanl.gov/abs/0901.1165}{{\tt arXiv:0901.1165}}.

\bibitem{Pshirkov:2011um}
M.~Pshirkov, P.~Tinyakov, P.~Kronberg, and K.~Newton-McGee, {\it {Deriving
  global structure of the Galactic Magnetic Field from Faraday Rotation
  Measures of extragalactic sources}},  {\em Astrophys.J.} {\bf 738} (2011)
  192, [\href{http://xxx.lanl.gov/abs/1103.0814}{{\tt arXiv:1103.0814}}].

\bibitem{Jansson:2012pc}
R.~Jansson and G.~R. Farrar, {\it {A New Model of the Galactic Magnetic
  Field}},  {\em Astrophys.J.} {\bf 757} (2012) 14,
  [\href{http://xxx.lanl.gov/abs/1204.3662}{{\tt arXiv:1204.3662}}].

\bibitem{Glennys_private}
G.~R. Farrar. private communication, 2012.

\bibitem{Han:2004aa}
J.-L. Han, K.~Ferriere, and R.~Manchester, {\it {The Spatial energy spectrum of
  magnetic fields in our galaxy}},  {\em Astrophys.J.} {\bf 610} (2004)
  820--826, [\href{http://xxx.lanl.gov/abs/astro-ph/0404221}{{\tt
  astro-ph/0404221}}].

\bibitem{DeMarco:2007eh}
D.~De~Marco, P.~Blasi, and T.~Stanev, {\it {Numerical propagation of high
  energy cosmic rays in the Galaxy. I. Technical issues}},  {\em JCAP} {\bf
  0706} (2007) 027, [\href{http://xxx.lanl.gov/abs/0705.1972}{{\tt
  arXiv:0705.1972}}].

\bibitem{Gleeson_1968ApJ}
L.~J. {Gleeson} and W.~I. {Axford}, {\it {Solar Modulation of Galactic Cosmic
  Rays}},  {\em \apj} {\bf 154} (Dec., 1968) 1011.

\bibitem{2011ApJ...735...83S}
R.~D. {Strauss}, M.~S. {Potgieter}, I.~{B{\"u}sching}, and A.~{Kopp}, {\it
  {Modeling the Modulation of Galactic and Jovian Electrons by Stochastic
  Processes}},  {\em \apj} {\bf 735} (July, 2011) 83.

\bibitem{2012Ap&SS.339..223S}
R.~D. {Strauss}, M.~S. {Potgieter}, I.~{B{\"u}sching}, and A.~{Kopp}, {\it
  {Modelling heliospheric current sheet drift in stochastic cosmic ray
  transport models}},  {\em \apss} {\bf 339} (June, 2012) 223--236.

\bibitem{1996ApJ...464..507C}
J.~M. {Clem}, D.~P. {Clements}, J.~{Esposito}, P.~{Evenson}, D.~{Huber},
  J.~{L'Heureux}, P.~{Meyer}, and C.~{Constantin}, {\it {Solar Modulation of
  Cosmic Electrons}},  {\em \apj} {\bf 464} (June, 1996) 507.

\bibitem{Longair}
M.~S. Longair, {\em High Energy Astrophysics}.
\newblock Cambridge University Press, 2010.

\bibitem{1988ApJ...334L...5G}
G.~{Ghisellini}, P.~W. {Guilbert}, and R.~{Svensson}, {\it {The synchrotron
  boiler}},  {\em \apjl} {\bf 334} (Nov., 1988) L5--L8.

\bibitem{Kamae:2006bf}
T.~Kamae, N.~Karlsson, T.~Mizuno, T.~Abe, and T.~Koi, {\it {Parameterization of
  Gamma, e+/- and Neutrino Spectra Produced by p-p Interaction in Astronomical
  Environment}},  {\em Astrophys.J.} {\bf 647} (2006) 692--708,
  [\href{http://xxx.lanl.gov/abs/astro-ph/0605581}{{\tt astro-ph/0605581}}].

\bibitem{Huang:2007wk}
C.-Y. Huang and M.~Pohl, {\it {Production of Neutrinos and Secondary Electrons
  in Cosmic Sources}},  {\em Astropart.Phys.} {\bf 29} (2008) 282--289,
  [\href{http://xxx.lanl.gov/abs/0711.2528}{{\tt arXiv:0711.2528}}].

\bibitem{Pohl_private}
M.~Pohl. private communication, 2012.

\bibitem{Oliveira-Costa_2008MNRAS}
A.~{de Oliveira-Costa}, M.~{Tegmark}, B.~M. {Gaensler}, J.~{Jonas}, T.~L.
  {Landecker}, and P.~{Reich}, {\it {A model of diffuse Galactic radio emission
  from 10 MHz to 100 GHz}},  {\em \mnras} {\bf 388} (July, 2008) 247--260,
  [\href{http://xxx.lanl.gov/abs/0802.1525}{{\tt arXiv:0802.1525}}].

\bibitem{Strong:2011wd}
A.~Strong, E.~Orlando, and T.~Jaffe, {\it {The interstellar cosmic-ray electron
  spectrum from synchrotron radiation and direct measurements}},  {\em
  Astron.Astrophys.} {\bf 534} (2011) A54,
  [\href{http://xxx.lanl.gov/abs/1108.4822}{{\tt arXiv:1108.4822}}].

\bibitem{Rybicki}
G.~B. Rybicki and A.~Lightman, {\em Radiative Processes in Astrophysics}.
\newblock Wiley-VCH, 2004.

\bibitem{Bringmann:2011py}
T.~Bringmann, F.~Donato, and R.~A. Lineros, {\it {Radio data and synchrotron
  emission in consistent cosmic ray models}},  {\em JCAP} {\bf 1201} (2012)
  049, [\href{http://xxx.lanl.gov/abs/1106.4821}{{\tt arXiv:1106.4821}}].

\bibitem{DiBernardo:2010is}
G.~Di~Bernardo, C.~Evoli, D.~Gaggero, D.~Grasso, L.~Maccione, et~al., {\it
  {Implications of the Cosmic Ray Electron Spectrum and Anisotropy measured
  with Fermi-LAT}},  {\em Astropart.Phys.} {\bf 34} (2011) 528--538,
  [\href{http://xxx.lanl.gov/abs/1010.0174}{{\tt arXiv:1010.0174}}].

\bibitem{1998ApJ...493..694M}
I.~V. {Moskalenko} and A.~W. {Strong}, {\it {Production and Propagation of
  Cosmic-Ray Positrons and Electrons}},  {\em \apj} {\bf 493} (Jan., 1998) 694,
  [\href{http://xxx.lanl.gov/abs/astro-ph/}{{\tt astro-ph/}}].

\bibitem{DellaTorre:2012zz}
S.~Della~Torre, P.~Bobik, M.~J. Boschini, C.~Consolandi, M.~Gervasi, et~al.,
  {\it {Effects of solar modulation on the cosmic ray positron fraction}},
  {\em Adv.Space Res.} {\bf 49} (2012) 1587--1592.

\bibitem{0004-637X-565-1-280}
I.~V. Moskalenko, A.~W. Strong, J.~F. Ormes, and M.~S. Potgieter, {\it
  Secondary antiprotons and propagation of cosmic rays in the galaxy and
  heliosphere},  {\em The Astrophysical Journal} {\bf 565} (2002), no.~1 280.

\bibitem{DiBernardo:2009ku}
G.~Di~Bernardo, C.~Evoli, D.~Gaggero, D.~Grasso, and L.~Maccione, {\it {Unified
  interpretation of cosmic-ray nuclei and antiproton recent measurements}},
  {\em Astropart.Phys.} {\bf 34} (2010) 274--283,
  [\href{http://xxx.lanl.gov/abs/0909.4548}{{\tt arXiv:0909.4548}}].

\bibitem{1996ARA&A..34..155B}
R.~{Beck}, A.~{Brandenburg}, D.~{Moss}, A.~{Shukurov}, and D.~{Sokoloff}, {\it
  {Galactic Magnetism: Recent Developments and Perspectives}},  {\em \araa}
  {\bf 34} (1996) 155--206.

\bibitem{Bulanov_1974Ap&SS}
S.~V. {Bulanov} and V.~A. {Dogel}, {\it {The Influence of the Energy Dependence
  of the Diffusion Coefficient on the Spectrum of the Electron Component of
  Cosmic Rays and the Radio Background Radiation of the Galaxy}},  {\em \apss}
  {\bf 29} (Aug., 1974) 305--318.

\bibitem{Delahaye:2010ji}
T.~Delahaye, J.~Lavalle, R.~Lineros, F.~Donato, and N.~Fornengo, {\it {Galactic
  electrons and positrons at the Earth:new estimate of the primary and
  secondary fluxes}},  {\em Astron.Astrophys.} {\bf 524} (2010) A51,
  [\href{http://xxx.lanl.gov/abs/1002.1910}{{\tt arXiv:1002.1910}}].

\bibitem{Strong:2010pr}
A.~Strong, T.~Porter, S.~Digel, G.~Johannesson, P.~Martin, et~al., {\it {Global
  cosmic-ray related luminosity and energy budget of the Milky Way}},  {\em
  Astrophys.J.} {\bf 722} (2010) L58--L63,
  [\href{http://xxx.lanl.gov/abs/1008.4330}{{\tt arXiv:1008.4330}}].

\bibitem{Webber_2008JGRA}
W.~R. {Webber} and P.~R. {Higbie}, {\it {Limits on the interstellar cosmic ray
  electron spectrum below \~{}1-2 GeV derived from the galactic polar radio
  spectrum and constrained by new Voyager 1 measurements}},  {\em Journal of
  Geophysical Research (Space Physics)} {\bf 113} (Nov., 2008) 11106.

\bibitem{Incagli:2010}
M.~{Incagli}, {\it {Astroparticle Physiscs with AMS02}},  in {\em American
  Institute of Physics Conference Series} (C.~{Cecchi}, S.~{Ciprini},
  P.~{Lubrano}, and G.~{Tosti}, eds.), vol.~1223 of {\em American Institute of
  Physics Conference Series}, pp.~43--49, Mar., 2010.

\bibitem{Torii:2008}
S.~{Torii} and {CALET Collaboration}, {\it {The CALET mission for detection of
  cosmic ray sources and dark matter}},  {\em Journal of Physics Conference
  Series} {\bf 120} (July, 2008) 062020.

\bibitem{Neronov:2011wi}
A.~Neronov, D.~Semikoz, and A.~Taylor, {\it {Low-energy break in the spectrum
  of Galactic cosmic rays}},  {\em Phys.Rev.Lett.} {\bf 108} (2012) 051105,
  [\href{http://xxx.lanl.gov/abs/1112.5541}{{\tt arXiv:1112.5541}}].

\bibitem{Dermer:2012bz}
C.~D. Dermer, {\it {Diffuse Galactic Gamma Rays from Shock-Accelerated Cosmic
  Rays}},  {\em Phys.Rev.Lett.} {\bf 109} (2012) 091101,
  [\href{http://xxx.lanl.gov/abs/1206.2899}{{\tt arXiv:1206.2899}}].

\bibitem{Kachelriess:2012fz}
M.~Kachelriess and S.~Ostapchenko, {\it {Deriving the cosmic ray spectrum from
  gamma-ray observations}},  {\em Phys.Rev.} {\bf D86} (2012) 043004,
  [\href{http://xxx.lanl.gov/abs/1206.4705}{{\tt arXiv:1206.4705}}].

\bibitem{Donato:2001eq}
F.~Donato, D.~Maurin, and R.~Taillet, {\it {Beta-radioactive cosmic rays in a
  diffusion model: test for a local bubble?}},  {\em Astron.Astrophys.} {\bf
  381} (2002) 539--559, [\href{http://xxx.lanl.gov/abs/astro-ph/0108079}{{\tt
  astro-ph/0108079}}].

\bibitem{2001ApJ...555..585M}
D.~{Maurin}, F.~{Donato}, R.~{Taillet}, and P.~{Salati}, {\it {Cosmic Rays
  below Z=30 in a Diffusion Model: New Constraints on Propagation Parameters}},
   {\em \apj} {\bf 555} (July, 2001) 585--596,
  [\href{http://xxx.lanl.gov/abs/astro-ph/}{{\tt astro-ph/}}].

\bibitem{Taillet:2002ub}
R.~Taillet and D.~Maurin, {\it {Spatial origin of galactic cosmic rays in
  diffusion models: 1. Standard sources in the galactic disk}},  {\em
  Astron.Astrophys.} {\bf 402} (2003) 971,
  [\href{http://xxx.lanl.gov/abs/astro-ph/0212112}{{\tt astro-ph/0212112}}].

\bibitem{Blasi:2012}
P.~{Blasi} and E.~{Amato}, {\it {Diffusive propagation of cosmic rays from
  supernova remnants in the Galaxy. II: anisotropy}},  {\em \jcap} {\bf 1}
  (Jan., 2012) 11, [\href{http://xxx.lanl.gov/abs/1105.4529}{{\tt
  arXiv:1105.4529}}].

\bibitem{Cholis:2011un}
I.~Cholis, M.~Tavakoli, C.~Evoli, L.~Maccione, and P.~Ullio, {\it {Diffuse
  Galactic Gamma Rays at intermediate and high latitudes. I. Constraints on the
  ISM properties}},  {\em JCAP} {\bf 1205} (2012) 004,
  [\href{http://xxx.lanl.gov/abs/1106.5073}{{\tt arXiv:1106.5073}}].

\bibitem{Evoli:2012ha}
C.~Evoli, D.~Gaggero, D.~Grasso, and L.~Maccione, {\it {A common solution to
  the cosmic ray anisotropy and gradient problems}},  {\em Phys.Rev.Lett.} {\bf
  108} (2012) 211102, [\href{http://xxx.lanl.gov/abs/1203.0570}{{\tt
  arXiv:1203.0570}}].

\bibitem{2005ApJ...622..759G}
K.~M. {G{\'o}rski}, E.~{Hivon}, A.~J. {Banday}, B.~D. {Wandelt}, F.~K.
  {Hansen}, M.~{Reinecke}, and M.~{Bartelmann}, {\it {HEALPix: A Framework for
  High-Resolution Discretization and Fast Analysis of Data Distributed on the
  Sphere}},  {\em \apj} {\bf 622} (Apr., 2005) 759--771,
  [\href{http://xxx.lanl.gov/abs/astro-ph/0409513}{{\tt astro-ph/0409513}}].

\end{thebibliography}\endgroup
\bibliographystyle{JHEP}

\end{document}